\documentclass[prd,reprint,superscriptaddress,nofootinbib]{revtex4-1}
\usepackage{amsmath}
\usepackage{amssymb}
\usepackage{graphicx}
\usepackage[caption=false]{subfig}
\usepackage{enumitem}
\usepackage{color}
\usepackage[percent]{overpic}
\usepackage{bm}
\usepackage{rotating}
\usepackage{hyperref}
\usepackage{pbox}
\usepackage{array}

\newcommand{\ket}[1]{{\left| {#1} \right>}}
\newcommand{\bra}[1]{{\left< {#1} \right|}}

\newcommand{\ii}{\mathrm{i}}

\DeclareMathOperator{\tr}{Tr}
\usepackage[normalem]{ulem}
\begin{document}

\title{Harvesting correlations from the quantum vacuum}

\author{Alejandro Pozas-Kerstjens}
\affiliation{Perimeter Institute for Theoretical Physics, 31 Caroline St N, Waterloo, Ontario, N2L 2Y5, Canada}
\author{Eduardo Mart\'{i}n-Mart\'{i}nez}
\affiliation{Perimeter Institute for Theoretical Physics, 31 Caroline St N, Waterloo, Ontario, N2L 2Y5, Canada}
\affiliation{Department of Applied Mathematics, University of Waterloo, Waterloo, Ontario, N2L 3G1, Canada}
\affiliation{Institute for Quantum Computing, University of Waterloo, Waterloo, Ontario, N2L 3G1, Canada}


\begin{abstract}
We analyze the harvesting of entanglement and classical correlations from the quantum vacuum to particle detectors. We assess the impact on the detectors' harvesting ability of the spacetime dimensionality, the suddenness of the detectors' switching, their physical size and their internal energy structure. Our study reveals several interesting dependences on these parameters that can be used to optimize the harvesting of classical and quantum correlations. Furthermore, we find that, contrary to previous belief, smooth switching is much more efficient than sudden switching in order to harvest vacuum entanglement, especially when the detectors remain spacelike separated. Additionally, we show that the reported phenomenology of spacelike entanglement harvesting is not altered by subleading-order perturbative corrections.
\end{abstract}

\maketitle

\section{Introduction}
It has been known for a long time that the vacuum state of a free quantum field contains correlations between time- and spacelike-separated regions \cite{Algebra1,Algebra2}. Besides the remarkable fundamental interest from the point of view of quantum foundations, the existence of this vacuum entanglement is a key ingredient in very interesting, recently-discovered phenomena such as Masahiro Hotta's quantum energy teleportation \cite{Hotta2009,HottaRev}. It is also at the core of long open problems such as the black hole information loss problem \cite{Preskill1993} and some of its proposed tentative solutions, such as the so-called ``black hole firewalls'', and black hole complementarity \cite{Susskind1993,Almheiri2012,Braunstein2013}.

A perhaps more surprising result is that this vacuum entanglement can be extracted from the field to Unruh-DeWitt particle detectors \cite{DeWitt1979} that couple to the field locally even when the two detectors are spacelike-separated, as pointed out, first by Valentini \cite{Valentini1991}, and later by Reznik \cite{Reznik2003,Reznik2005}. This phenomenon has become known as \textit{entanglement harvesting} \cite{Salton2015}.

Since Unruh-DeWitt detectors can, in some regimes, be a good approximation to the light-matter interaction \cite{Martin-Martinez2013,Alvaro}, these pioneering results may imply that it is possible to extract entanglement from the electromagnetic vacuum to atomic qubits, where it could be used as a resource, although, for this purpose, one has to be careful with the impact of time synchronization on entanglement harvesting \cite{Barry}. Indeed, it has been proved that it is possible to devise quantum optical setups where entanglement can be sustainably and reliably extracted from a quantum field and distilled into Bell pairs, that can be later used as a resource for quantum information tasks. This technique is known as \textit{entanglement farming} \cite{Farming}. Moreover, there have been several exploratory works on the experimental feasibility of timelike and spacelike entanglement harvesting in atomic physics and superconducting circuits \cite{RalphPastFut1,RalphPastFut2,PastFutPRL}.

Entanglement harvesting was proven, by Ver Steeg and Menicucci, to be sensitive to the structure of the background spacetime in which it is performed \cite{VerSteeg2009}. In particular, they proved that entanglement harvesting can distinguish between a thermal background and the Gibbons-Hawking radiation background of an expanding universe \cite{VerSteeg2009,cosmo1,cosmo2}, and it is also sensitive to the topology of spacetime \cite{Martin-MartinezSmithTerno}.

Not only that, entanglement harvesting has been proven to be very sensitive to the state of motion of the detectors, and the boundary conditions on the field on which is performed. This has led to proposals of applications in metrology such as range-finding \cite{Salton2015} and even quantum seismology \cite{QSeismo}.

As a fundamental phenomenon, entanglement harvesting is therefore relatively well understood. However, little is still known about how it is affected by (and possibly optimized over) variations of the specific parameters of the setup, such as how fast the detectors are switched on, the dimension of spacetime, the physical size of the detectors or the nature of their internal degrees of freedom. This is particularly important in the case of spacelike entanglement harvesting, which would constitute a direct proof of the existence of vacuum correlations. We can find several hypotheses and intuitions about some of these aspects in the literature. For example, in one of the original papers by Reznik \textit{et al.} \cite{Reznik2005}, in which they maximize entanglement harvesting using very fast-varying super-oscillatory switching functions, or in \cite{Brown2012} (in the context of harmonic oscillator-based non-perturbative methods for particle detectors \cite{Brown2012,Bruschi2012}), where it was speculated that a sudden switching might be more efficient than a smooth one to harvest entanglement.

In this paper we present a thorough study of both entanglement harvesting and the harvesting of classical correlations, and how they are affected by the dimension of spacetime, the physical size of the detectors, their internal energy structure and the smoothness of the switching of their coupling to the field. Remarkably, and contrary to previous belief, we find that smoother switchings are much more efficient than sudden switchings in order to harvest vacuum entanglement. Namely, we find that while for a smooth Gaussian switching it is always possible to choose detector setups that allow for spacelike entanglement harvesting, this is not the case for sudden switchings. We trace back this result to the fact that sudden switchings increase the amount of local noise that the particle detectors experience \cite{Louko2008}, which hinders their ability to harvest vacuum entanglement.

We show that entanglement harvesting is rather insensitive to the dimensionality of spacetime, but this is not the case for the harvesting of classical correlations. Namely, for a 3+1-dimensional spacetime, mutual information is more efficiently harvested from the vacuum than in 1+1 dimensions when the detectors are in lightlike contact. On the other hand, reducing the dimensionality of spacetime improves the ability of the detectors to harvest correlations when they are spacelike separated.

We also show that finite-size, but small detectors (as compared to their interaction time with the field) do not behave in a fundamentally different way to pointlike detectors, in the regimes where the pointlike approximation is not ill defined. When the size of the detectors is increased and it becomes comparable to their interaction times, larger detectors are much less efficient to harvest entanglement than smaller ones.

As for the dependence on the detectors' energy gap, we show that the situation is radically different in the cases of sudden and smooth switching. For the latter it is always possible to tune the detectors' energy gap in order to harvest spacelike entanglement for a given setup, whereas for a very fast switching it is generally not possible to do so.

Finally, we have also analyzed vacuum entanglement harvesting at higher orders in perturbation theory, showing that going beyond leading order does not reveal new phenomenology. Therefore a leading-order perturbative approximation is generally enough to identify the regimes in which Unruh-DeWitt detectors can harvest quantum entanglement from the field vacuum.

\section{Setup}\label{sec:general}

We will model two particle detectors (A and B) with the well-known Unruh-DeWitt model \cite{DeWitt1979}. Although simple, this detector model comprises most of the fundamental features of the light-matter interaction when there is no exchange of angular momentum \cite{Martin-Martinez2013,Alvaro}. The Unruh-DeWitt detectors (from now on referred to as the `atoms' or `detectors') interact with a background scalar field via the following Hamiltonian in the interaction picture,
\begin{equation}
H_I(t)=\!\!\!\!\!\sum_{\nu\in\{\text{A,\,B}\}}\!\!\!\!\lambda_{\nu}\chi_{\nu}(t)\mu_{\nu}(t)\!\int\text{d}^n\bm{x}\,F_{\nu}(\bm{x}-\bm{x}_{\nu})\phi(\bm{x},t),
\label{eq:hamiltonian}
\end{equation}
where $\lambda_{\nu}$ is the overall coupling strength, $\mu_{\nu}(t)$ is the monopole moment of each detector
\begin{equation}
\mu_{\nu}(t)=\sigma^+_{\nu}e^{\ii\Omega_{\nu}t}+\sigma^-_{\nu}e^{-\ii\Omega_{\nu}t}\label{eq:moment},
\end{equation}
($\sigma^\pm_\nu$ are SU(2) ladder operators), $F_\nu(\bm{x})$ are the spatial smearing functions of each detector and $\bm x_\nu$  their respective center-of-mass positions. $\chi_\nu(t)$ is detector $\nu$'s switching function, which controls the interaction time and the coupling strength of each atom with the field. For our purposes we are going to consider switching functions that are strongly suppressed outside of finite time intervals (so as to have finite-duration interactions).

Typically, if we think of the UDW model as a model of the light-matter interaction, the spatial support of the atom can be associated with the spatial probability profile of the atomic wavefunctions \cite{Alvaro}.
In an $n$+1-dimensional flat spacetime, the scalar field can be expanded in terms of plane-wave modes in the following way
\begin{equation}
\phi(\bm{x},t)=\int\frac{\text{d}^n\bm{k}}{\sqrt{(2\pi)^n2|\bm{k}|}}
\left[a^\dagger_{\bm{k}}e^{\ii(|\bm{k}|t-\bm{k}\cdot\bm{x})}+\text{H.c.}\right].\label{eq:field}
\end{equation}
The creation and annihilation operators $a_{\bm k}$ and $a_{\bm k}^\dagger$ satisfy canonical commutation relations  $[a_{\bm k},a_{\bm k'}^\dagger]=\delta^{(n)}(\bm k-\bm k')$.

The integral with respect to $\bm{x}$ can be easily performed yielding the Fourier transform of the spatial profile.
\begin{align}
H_I&=\sum_{\nu}\lambda_{\nu}\chi_{\nu}(t)\mu_{\nu}(t)\int\frac{\text{d}^n\bm{k}}{\sqrt{2|\bm{k}|}} \\
&\times\left[a_{\bm{k}}e^{-\ii(|\bm{k}|t-\bm{k}\cdot\bm{x}_{\nu})}\tilde{F}_{\nu}(\bm{k})+a^{\dagger}_{\bm{k}}e^{\ii(|\bm{k}|t-\bm{k}\cdot\bm{x}_{\nu})}\tilde{F}_{\nu}(\bm{-k})\right],\nonumber
\end{align}
where the form factor of the atoms is defined as
\begin{equation}
\tilde{F}_{\nu}(\bm{k})=\frac{1}{\sqrt{(2\pi)^n}}\int\text{d}^n\bm{x}\,F_{\nu}(\bm{x})e^{\ii\bm{k}\cdot\bm{x}}.
\end{equation}

From now on, we will consider that both detectors will have the same shape and their spatial profile will be a real function, therefore $\tilde{F}_{\nu}(\bm{k})=\tilde{F}_{\nu}(-\bm{k})=\tilde{F}(\bm{k})$.

The time evolution generated by \eqref{eq:hamiltonian} can be obtained perturbatively through a Dyson expansion of the time evolution operator:
\begin{align}
U=\openone\underbrace{-\ii\int_{-\infty}^{\infty}\!\!\!\text{d}t\,H(t)}_{U^{(1)}}\underbrace{-\!\!\int_{-\infty}^{\infty}\!\!\!\!\text{d}t\int_{-\infty}^{t}\!\!\!\!\!\!\text{d}t^{\prime}\,H(t)H(t^{\prime})}_{U^{(2)}}+\dots\label{eq:evo}
\end{align}

If the initial state of the detectors-field system is $\rho_0$, the evolved state  will be given by $\rho=U\rho_0U^{\dagger}$. Let us define the notation $\mathcal{O}\left(\lambda_\nu^n\right)$ to represent terms that are proportional to $\lambda_A^k\lambda_B^l$ $\forall\,k,l$ such that $k+l=n$. We will denote the $\mathcal{O}\left(\lambda_\nu^{i+j}\right)$ as a sum of terms of the form perturbative contributions to the time-evolved density matrix as
\begin{equation}
\rho^{(i,j)}=U^{(i)}\rho_0U^{(j)\dagger}.\label{rhoij}
\end{equation}

Therefore we can write the time-evolved density matrix as a sum of terms of the form of (\ref{rhoij}):
\begin{equation}
\rho=\rho_0+\rho^{(1,0)}+\rho^{(0,1)}+\rho^{(2,0)}+\rho^{(0,2)}+\rho^{(1,1)}+\rho^{(1,2)}+\dots
\end{equation}

Since we are going to analyze entanglement and correlations harvesting from the vacuum, we consider that the initial state of the detectors-field system is
\begin{equation}
\rho_0=\ket{0}\bra{0}\otimes\rho_{AB,0},
\end{equation}
where $\ket0$ is the vacuum state of the scalar field and $\rho_{AB,0}$ is the initial state of the detectors. We will be interested in the partial state of the detectors after their interaction with the field, which is given by
\begin{equation}\label{tracedd}
\rho_{AB}=\tr_\phi(U\rho_0U^{\dagger}).
\end{equation}
This means that the non-diagonal terms in the field produced by time evolution will be of no relevance for our purposes. In particular, any contribution $\rho^{(i,j)}$ for which the parities of $i$ and $j$ are different (i.e, for all the contributions with odd powers of the coupling strength $\lambda_\nu$) will give a zero contribution to the detectors' final state \eqref{tracedd}, as long as the initial state of the field  is diagonal in the Fock basis (as it is the case of the vacuum).

Consequently, in the perturbative expansion of $\rho_{AB}$, the first order correction is trivially zero for the reason explained above. To leading order in the coupling strength, the two detectors' time-evolved density matrix is given by
\begin{equation}
\rho_{AB}=\rho_{AB,0}+\rho_{AB}^{(2,0)}+\rho_{AB}^{(0,2)}+\rho_{AB}^{(1,1)}+\mathcal{O}(\lambda_\nu^4),\label{rhoAB}
\end{equation}
where $\rho_{AB}^{(i,j)}=\text{Tr}_\phi\left[\rho^{(i,j)}\right]$.

We are going to consider the case in which both detectors are in their respective ground states
\begin{equation}
\rho_{AB,0}=\ket{g_A}\bra{g_A}\otimes\ket{g_B}\bra{g_B}.
\end{equation}

From \eqref{rhoAB}, $\rho_{AB}$ takes the following matrix representation
\begin{equation}
\rho_{\text{AB}}=\begin{pmatrix}
1-\mathcal{L}_{AA}-\mathcal{L}_{BB} & 0 & 0 & \mathcal{M}^* \\
0 & \mathcal{L}_{AA} & \mathcal{L}_{AB} & 0 \\
0 & \mathcal{L}_{BA} & \mathcal{L}_{BB} & 0 \\
\mathcal{M} & 0 & 0 & 0
\end{pmatrix}
+\mathcal{O}(\lambda_{\nu}^4) \label{eq:seconddens}
\end{equation}
in the basis 
\begin{equation}
\left\{\ket{g_A}\otimes\ket{g_B},\,\ket{e_A}\otimes\ket{g_B},\,\ket{g_A}\otimes\ket{e_B},\,\ket{e_A}\otimes\ket{e_B}\right\}.
\end{equation}

The explicit expressions of $\mathcal{L}_{\mu\nu}$ and $\mathcal{M}$ can be easily obtained from Eq. \eqref{eq:evo} after substituting \eqref{eq:hamiltonian}, \eqref{eq:moment} and \eqref{eq:field}, yielding
\begin{align}
\mathcal{L}_{\mu\nu}&=\int\text{d}^n\bm{k}\,L_\mu(\bm{k})L_\nu(\bm{k})^*,\label{Lmunu}\\
\mathcal{M}&=\int\text{d}^n\bm{k}\,M(\bm{k}),\label{M}
\end{align}
where $L_\mu(\bm{k})$ and $M(\bm{k})$ are~\footnote{A note for those repeating this calculation: to obtain~Eq.~\eqref{Mnotint} from \eqref{eq:evo} we performed a change of variables in the integral over momentum $\bm{k}\rightarrow-\bm{k}$ for one of the summands that leaves the integral invariant but allows us to write a more compact expression with the same phase $e^{\ii\bm{k}\cdot(\bm{x}_{A}-\bm{x}_{B})}$ for both summands.}
\begin{align}
L_{\mu}(\bm{k})=&\lambda_{\mu}\frac{e^{-\ii\bm{k}\cdot\bm{x}_{\mu}}\tilde{F}(\bm{k})}{\sqrt{2|\bm{k}|}}\int_{-\infty}^{\infty}\text{d}t_1\,\chi_{\mu}(t_1)e^{\ii(|\bm{k}|+\Omega_{\mu})t_1},\label{Lmunotint}\\
M(\bm{k})=&-\lambda_{A}\lambda_{B}e^{\ii\bm{k}\cdot(\bm{x}_{A}-\bm{x}_{B})}\frac{[\tilde{F}(\bm{k})]^2}{2|\bm{k}|}\notag\\
&\times\int_{-\infty}^{\infty}\text{d}t_1\int_{-\infty}^{t_1}\text{d}t_2\,e^{-\ii|\bm{k}|(t_1-t_2)}\notag\\
&\left[\chi_{A}(t_1)\chi_{B}(t_2)e^{\ii(\Omega_{A} t_1+\Omega_{B} t_2)}\right.\notag\\
&\left.+\chi_{B}(t_1)\chi_{A}(t_2)e^{\ii(\Omega_{B} t_1+\Omega_{A} t_2)}\right]. \label{Mnotint}
\end{align}

The expressions above are rather general and can be easily particularized to any switching and spatial profiles for any dimension.

We would like to analyze under which set of general conditions it is possible to harvest classical correlations and entanglement from the field to the detectors. With this aim, we will consider different switching modalities (sudden versus Gaussian), different characteristic detector sizes (pointlike versus non-negligible Gaussian smearing), different spacetime dimensions (1+1-dimensional ---as in long wave guides or optical fibers-- versus 3+1-dimensional ---as in free space--) and a range of different detector internal energy scales.

The first step is to evaluate the integrals \eqref{Lmunotint} and \eqref{Mnotint} for the different cases that we will consider. Let us first focus on the spatial profile. We will choose the following Gaussian smearing
\begin{equation}
F(\bm{x})=\frac{1}{(\sqrt{\pi}\sigma)^n}e^{-\bm{x}^2/\sigma^2}
\end{equation}
which in turn enters equations \eqref{Lmunotint} and \eqref{Mnotint} via its Fourier transform
\begin{equation}
\tilde{F}(\bm{k})=\frac{1}{\sqrt{(2\pi)^n}}e^{-\frac{1}{4}|\bm{k}|^2\sigma^2}.\label{smear}
\end{equation}
It will be relevant to consider the limit where the detector is pointlike localized in space ($\sigma\rightarrow0$). Namely
\begin{equation}
F(\bm{x})=\delta^{(n)}(\bm{x})\Rightarrow\tilde{F}(\bm{k})=\frac{1}{\sqrt{\left(2\pi\right)^n}}.\label{point}
\end{equation}

\subsection{3+1 dimensions\label{3D}}

We will first consider the case of three spatial dimensions. We will study the following scenarios
\begin{enumerate}
\item[A1.] Gaussian switching and Gaussian spatial smearing.
\item[A2.] Gaussian switching and pointlike detectors.
\item[A3.] Sudden switching and Gaussian spatial smearing.
\item[A4.] Sudden switching and (almost) pointlike detectors.
\end{enumerate}
\subsubsection{Gaussian switching functions and Gaussian smearing\label{3DGauss}}
Let us first consider the case in which the detectors are turned on in a smooth manner, following a Gaussian profile
\begin{equation}
\chi_\nu(t)=e^{-(t-t_\nu)^2/T^2}.\label{gauss}
\end{equation}

With this switching function, all time integrals in \eqref{Lmunotint} and \eqref{Mnotint} admit analytic closed forms. We will assume that both detectors have the same energy gap $\Omega\equiv\Omega_A=\Omega_B$ and that both couple to the field with the same strength $\lambda\equiv\lambda_A=\lambda_B$. This allows us to scale all the parameters in the system relative to the characteristic timescale of the switching function $T$, as suggested in \cite{VerSteeg2009,Salton2015}. Defining the dimensionless magnitudes $\alpha=\Omega T$, $\bm\beta_\mu=\bm{x}_\mu/T$, $\delta=\sigma/T$, $\bm{\kappa}=\bm{k}T$ and $\tau_\mu=t_\mu/T$  (a summary of all the dimensionless parameters used throughout the paper can be found in Table \ref{notation}), \eqref{Lmunotint} and \eqref{Mnotint} can be recast as
\begin{align}
L_{\mu}(\bm\kappa)=&\lambda \,T^{1/2}\,\frac{e^{-\ii\bm{\kappa}\cdot\bm{\beta}_{\mu}}e^{-\frac{1}{4}\bm{\kappa}^2\delta^2}}{\sqrt{2|\bm{\kappa}|\left(2\pi\right)^3}}G_1(\bm{\kappa},\tau_\mu)\label{LG},\\
M(\bm{\kappa})=&-\lambda^2\,T\,e^{\ii\bm{\kappa}\cdot(\bm{\beta}_{A}-\bm{\beta}_{B})}\frac{e^{-\frac{1}{2}\bm{\kappa}^2\delta^2}}{2|\bm{\kappa}|\left(2\pi\right)^3}G_2(\bm{\kappa})\label{MG}.
\end{align}
where, taking $\tau_i=(t_i-t_A)/T$ as dimensionless integration variables, we can write
\begin{align}
&G_1(\bm{\kappa},\tau_\mu)=T\int_{-\infty}^\infty\text{d}\tau_1\,e^{-\left(\tau_1-\tau_\mu\right)^2}e^{\ii(|\bm{\kappa}|+\alpha)\tau_1},\label{G1}\\
&G_2(\bm{\kappa})=e^{2\ii\alpha\tau_A}T^2\int_{-\infty}^\infty\!\!\!\text{d}\tau_1\int_{-\infty}^{\tau_1}\!\!\!\text{d}\tau_2\,e^{\ii\alpha(\tau_1+\tau_2)}e^{-\ii|\bm{\kappa}|(\tau_1-\tau_2)}\notag\\
&\qquad\qquad\times\left(e^{-\left(\tau_1-\gamma\right)^2}e^{-\tau_2^2}+e^{-\left(\tau_2-\gamma\right)^2}e^{-\tau_1^2}\right),\label{G2}
\end{align}
where $\gamma=\left(t_B-t_A\right)/T$ is the normalized separation between the switching functions' centers.

\begin{table} 
\begin{ruledtabular}
{\setlength{\extrarowheight}{5pt}\begin{tabular}{ccl}
\pbox{10cm}{Dimensionless\\variable} & Expression & Physical meaning\\[5pt]
\hline & &\\[-13pt]
\hline$\alpha$ & $\Omega T$  & Energy gap\\[5pt]
\hline$\bm\beta_\mu$ & $\bm x_\mu/T$ & Detectors' positions\\[5pt]
\hline$\beta$ & $d/T$ & Spatial distance\\[5pt]
\hline$\gamma$ & $\Delta/T$ & Time delay\\[5pt]
\hline$\delta$ & $\sigma /T$ & Detectors' size\\[5pt]
\hline$\bm\kappa$, $\bm\eta$ & $\bm k T$, $\bm q T$ & Momenta\\[5pt]
\hline$\tau$ & $t/ T$ & Time parameter \\
\end{tabular}}
\caption{Collection of all the dimensionless quantities that are used throughout this paper. Notice that $d=|\bm x_B - \bm x_A|$, $\Delta=t_B-t_A$ and $\beta=|\bm\beta_B-\bm\beta_A|$.}\label{notation}
\end{ruledtabular}
\end{table}

The integral \eqref{G1} can be readily analytically evaluated. Obtaining a closed form for \eqref{G2} is more involved, but it can be accomplished via parametric differentiation under the integral sign and solving the resulting differential equation (see details in Appendix \ref{AppendixGaussian}). The result in both cases is
\begin{align}
\label{bothgausstime}
G_1(\bm{\kappa},\tau_\mu)=&\sqrt{\pi }\,T\,e^{-\frac{1}{4} \left(|\bm{\kappa}|+\alpha \right)^2}e^{\ii(|\bm{\kappa}|+\alpha )\tau_\mu},\\
G_2(\bm{\kappa})=&\, \frac{\pi}{2}T^2e^{2\ii\alpha\tau_A}e^{-\frac{1}{2} \left(\alpha ^2+\bm\kappa ^2-2\ii\gamma\alpha\right)}\nonumber\\
&\times\left[E(\bm\kappa,\gamma)+E(\bm\kappa,-\gamma)\right]\nonumber.
\end{align}
where, for simplicity, we define
\begin{equation}
E(\bm\kappa,\gamma)=e^{\ii \gamma|\bm\kappa|}\left[1-\text{erf}\left(\frac{\gamma +\ii |\bm\kappa| }{\sqrt{2}}\right)\right].
\end{equation}

Using \eqref{LG}, \eqref{MG} and \eqref{bothgausstime} we can evaluate \eqref{Lmunu} and \eqref{M}, yielding (see Appendix \ref{AppendixGaussian})
\begin{align}
\nonumber \mathcal{L}_{AA}\!=&\frac{\lambda^2}{4\pi^2 T^2}\!\!\int_0^\infty\!\!\!\!\text{d}|\bm{\kappa}|\,|\bm\kappa|\,e^{-\frac{1}{2}\bm{\kappa}^2\delta^2}G_1(\bm\kappa,\tau_A)G_1^*(\bm\kappa,\tau_A)\notag\\
=&\frac{\lambda ^2 e^{-\frac{1}{2}\alpha^2}}{8 \pi \left(1+\delta ^2\right)} \!\!\left[2-\frac{\sqrt{2 \pi }\, \alpha \,e^{\frac{\alpha ^2}{2 \left(1+\delta ^2\right)}} \text{erfc}\left(\frac{\alpha }{\sqrt{2} \sqrt{1+\delta ^2}}\right)}{\sqrt{1+\delta ^2}}\right]\!,\label{LAAGS3D}
\end{align}
\begin{align}
\mathcal{L}_{AB}\!=&\frac{\lambda^2}{4\pi^2T^2\beta}\!\!\int_0^\infty\!\!\!\!\text{d}|\bm{\kappa}|\,\sin(|\bm{\kappa}|\beta)e^{-\frac{1}{2}\bm{\kappa}^2\delta^2}G_1(\bm\kappa,0)G_1^*(\bm\kappa,\gamma)\notag\\
=&\frac{\ii\lambda^2 e^{-\frac{1}{2} \alpha^2}e^{-\ii\alpha\gamma}}{8 \sqrt{2\pi} \beta\sqrt{1+\delta^2}}\notag\\
&\times\left\{e^{-\frac{\left(\beta+\gamma-\ii \alpha \right)^2}{2 \left(1+\delta^2\right)}} \text{erfc}\left(\ii\frac{\beta+\gamma-\ii \alpha }{\sqrt{2} \sqrt{1+\delta^2}}\right)\right.\notag\\
&\left.- e^{-\frac{\left(\beta-\gamma+\ii \alpha \right)^2}{2 \left(1+\delta^2\right)}} \text{erfc}\left(-\ii\frac{\beta-\gamma+\ii\alpha }{\sqrt{2} \sqrt{1+\delta^2}}\right)\right\},\label{LABGS3D}
\end{align}
\begin{align}
|\mathcal{M}|=&\frac{\lambda^2}{4\pi^2T^2\beta}\Bigg|\int_0^{\infty}\text{d}|\bm{\kappa}|\,\sin (\beta  |\bm{\kappa}| )\,e^{-\frac{1}{2} \delta ^2 \bm{\kappa} ^2}G_2(\bm{\kappa})\Bigg|\notag\\
&=\frac{\lambda^2e^{-\frac{1}{2}\alpha ^2}}{8\pi\beta}\bigg|\int_0^{\infty}\text{d}|\bm{\kappa}|\,\sin (\beta |\bm{\kappa}| )\,e^{-\frac{1}{2} \left(1+\delta ^2\right) \bm{\kappa} ^2}\notag\\
&\times\left[E(\bm\kappa,\gamma)+E(\bm\kappa,-\gamma)\right]\Big|,\label{MGS3D}
\end{align}
where we have defined $\beta=|\bm\beta_B-\bm\beta_A|=d/T$ where \mbox{$d=|\bm x_B-\bm x_A|$} is the distance between the detectors' centers of mass, and $\text{erfc}(x)=1-\text{erf}(x)$ is the complementary error function. Note that, with the assumptions made, $\mathcal{L}_{BB}=\mathcal{L}_{AA}$.

This expression is general for any separation between the centers of the Gaussians $\gamma=\left(t_B-t_A\right)/T$. We note that the integral over $|\bm\kappa|$ admits an approximate analytic closed form when $\gamma$ is large enough to neglect the overlap between the two Gaussian switchings. To see this, we start from \eqref{Mnotint} and notice that $\chi_\nu(t)$ are Gaussian functions with standard deviation $s_T=T/\sqrt{2}$. When the centers of the two Gaussians are separated by $\Delta=t_B-t_A\geq 7T/\sqrt{2}$ (i.e. more than seven times the standard deviation), the two Gaussians effectively do not overlap. Note that, for $\Delta\geq 7T/\sqrt2$, the overlap between the Gaussian switchings is suppressed by a factor smaller than $e^{-49/2}\sim 10^{-11}$, this is, any effect of that overlap will be suppressed at least $10^{-11}$ times as compared to the contributions of the vicinity of the Gaussian maxima, rendering that overlap region completely negligible for our purposes. Assuming without loss of generality that the detector A is switched on before the detector B, in this case the first contribution of \eqref{Mnotint} is very approximately zero, and $|\mathcal{M}|\approx|\mathcal{M}_\text{non}|$, where
\begin{align}
\left|\mathcal{M}_{\text{non}}\right|
=&\frac{\lambda^2e^{-\frac{1}{2}\alpha^2}}{8 \sqrt{2 \pi} \beta \sqrt{1+\delta^2}}\notag\\
&\times  \left|e^{-\frac{(\beta-\gamma)^2}{2 \left(1+\delta^2\right)}} \left[1+ \text{erf}\left(\ii\frac{\beta-\gamma}{\sqrt{2} \sqrt{1+\delta^2}}\right)\right]\right.\notag\\
&\left.-e^{-\frac{(\beta+\gamma)^2}{2 \left(1+\delta^2\right)}} \left[1-\text{erf}\left(\ii\frac{\beta+\gamma}{\sqrt{2} \sqrt{1+\delta^2}}\right)\right]\right|.\label{MGS3Dnonover}
\end{align}
\subsubsection{Gaussian switching functions and pointlike detectors}
For pointlike detectors we take the limit $\delta\rightarrow 0$ in Eqs. \eqref{LAAGS3D}, \eqref{LABGS3D}, \eqref{MGS3D} and \eqref{MGS3Dnonover}, yielding
\begin{align}
\mathcal{L}_{AA}=&\frac{\lambda^2 e^{-\frac{1}{2} \alpha^2}}{8 \pi}\left[2-e^{\frac{1}{2}\alpha^2}\sqrt{2\pi}\,\alpha\, \text{erfc}\left(\frac{\alpha}{\sqrt{2}}\right)\right],
\end{align}
\begin{align}
\mathcal{L}_{AB}=&\frac{\ii\lambda^2 e^{-\frac{1}{2} \alpha^2}e^{-\ii\alpha\gamma}}{8 \sqrt{2\pi} \beta}\notag\\
&\times\left\{e^{-\frac{\left(\beta+\gamma-\ii \alpha \right)^2}{2}} \text{erfc}\left(\ii\frac{\beta+\gamma-\ii \alpha }{\sqrt{2}}\right)\right.\notag\\
&\left.- e^{-\frac{\left(\beta-\gamma+\ii \alpha \right)^2}{2}} \text{erfc}\left(-\ii\frac{\beta-\gamma+\ii\alpha }{\sqrt{2}}\right)\right\},
\end{align}
\begin{align}
|\mathcal{M}|=&\frac{\lambda^2e^{-\frac{1}{2}\alpha ^2}}{8\pi\beta}\Bigg|\int_0^{\infty}\text{d}|\bm{\kappa}|\,\sin (\beta  |\bm{\kappa}| )\,e^{-\frac{1}{2} \bm{\kappa} ^2}\notag\\
&\times\left[E(\bm\kappa,\gamma)+E(\bm\kappa,-\gamma)\right]\Bigg|,
\end{align}
\begin{align}
\left|\mathcal{M}_{\text{non}}\right|=&\frac{\lambda^2e^{-\frac{1}{2}\alpha^2}}{8 \sqrt{2 \pi} \beta}\notag\\
&\times  \left|e^{-\frac{(\beta-\gamma)^2}{2}} \left[1+\text{erf}\left(\ii\frac{\beta-\gamma}{\sqrt{2} }\right)\right]\right.\notag\\
&\left.-e^{-\frac{(\beta+\gamma)^2}{2}} \left[1-\text{erf}\left(\ii\frac{\beta+\gamma}{\sqrt{2}}\right)\right]\right|.
\end{align}

It is worth mentioning that in this case there is another situation in which $\mathcal{M}$ has an analytic closed form, which is when the two identical detectors' switching functions are in perfect overlap (i.e, $\gamma=0$). In this case \cite{Martin-MartinezSmithTerno},
\begin{equation}
\mathcal{M}_{\text{coinc}}=\frac{\lambda^2e^{-\frac{1}{2} \left(\alpha ^2+\beta ^2\right)}}{4 \sqrt{2 \pi } \beta } \left[ \text{erfi}\left(\frac{\beta }{\sqrt{2}}\right)-\ii\right],
\end{equation}
where $\text{erfi}(x)=-\ii\,\text{erf}(\ii x)$ is the imaginary error function.

\subsubsection{Sudden switching functions and Gaussian smearing\label{3DSS}}
Let us now consider that the detector $\nu$ is switched on in an abrupt manner at $T_\nu^\text{on}$ and switched off in the same way at $T_\nu^\text{off}$. Namely, the switching functions in \eqref{Lmunotint} and \eqref{Mnotint} are now 
\begin{equation}
\chi_\nu(t)=
\begin{cases}
1 &\mbox{\text{if} } T_\nu^{\text{on}}<t<T_\nu^{\text{off}}, \\
0 & \mbox{\text{otherwise}}.
\end{cases}\label{suddenswitching}
\end{equation}

For this switching function, the time integrals also admit closed-form expressions. From \eqref{Lmunotint} and \eqref{Mnotint} we get
\begin{align}
L_{\mu}(\bm{\kappa})=&\lambda \,T^{1/2}\, \frac{e^{-\ii\bm{\kappa}\cdot\bm{\beta}_{\mu}}e^{-\frac{1}{4}\bm{\kappa}^2\delta^2}}{\sqrt{2|\bm{\kappa}|\left(2\pi\right)^3}}S_1(\bm{\kappa},\tau_\mu)\label{LS},\\
M(\bm{\kappa})=&-\lambda^2\,T\,e^{\ii\bm{\kappa}\cdot(\bm{\beta}_{A}-\bm{\beta}_{B})}\frac{e^{-\frac{1}{2}\bm{\kappa}^2\delta^2}}{2|\bm{\kappa}|\left(2\pi\right)^3}S_2(\bm{\kappa})\label{MS},
\end{align}
where, similar to the Gaussian case, $T=T_\nu^{\text{off}}-T_\nu^{\text{on}}$ is the timescale of the interaction, and $S_1(\bm\kappa,\tau_\mu)$ and $S_2(\bm\kappa)$ are defined by
\begin{align}
S_1(\bm{\kappa},\tau_\mu)&=T\int_{\tau_\mu^\text{on}}^{\tau_\mu^\text{off}}\text{d}\tau_1\,e^{\ii(|\bm{\kappa}|+\alpha)\tau_1}\label{S1}\\
S_2(\bm{\kappa})&=\int_{-\infty}^{\infty}\text{d}t_1\int_{-\infty}^{t_1}\text{d}t_2\,e^{\ii\Omega(t_1+t_2)}e^{-\ii|\bm{k}|(t_1-t_2)}\notag\\
&\times\left[\chi_A(t_1)\chi_B(t_2)+\chi_B(t_1)\chi_A(t_2)\right].\label{S2}
\end{align}

As in the Gaussian case, \eqref{S1} can be computed in a straightforward manner, while the case of \eqref{S2} needs a careful analysis (see Appendix \ref{AppendixSudden}). These integrals yield
\begin{align}\label{S1int}
S_1(\bm\kappa,\tau_\mu)=&-\frac{\ii T \left(e^{\ii \tau_\mu^\text{off} (|\bm{\kappa}|+\alpha )}-e^{\ii \tau_\mu^\text{on} (|\bm{\kappa}|+\alpha )}\right)}{|\bm{\kappa}|+\alpha },
\end{align}
and $S_2(\kappa)$ simplifies to
\begin{align}\label{S2nonint}
{S_2}(\bm\kappa)&={S_2}_\text{non}(\bm\kappa)\\
&=T^2e^{\ii\gamma(\alpha-|\bm\kappa|)}\frac{\left(e^{\ii (\alpha -\left| \bm{\kappa} \right| )}-1\right)\left(e^{\ii \left(\alpha +\left| \bm{\kappa} \right| \right)}-1\right)}{\bm{\kappa}^2-\alpha ^2}\nonumber
\end{align}
when $T_B^\text{on}>T_A^\text{off}$  (no overlap between the switching functions), and to
\begin{align}\label{S2overint}
{S_2}(\bm\kappa)&={S_2}_\text{over}(\bm\kappa)\\
&=\frac{T^2}{\alpha^2-\bm\kappa^2}\left[e^{\ii(\gamma+1)(\alpha-|\bm\kappa|)}\left( e^{2\ii\gamma|\bm\kappa|}-e^{\ii(\alpha+|\bm\kappa|)}\right)\right.\notag\\
&\left.+e^{\ii(\gamma+1)(\alpha-|\bm\kappa|)}-e^{\ii\gamma(\alpha-|\bm\kappa|)}-\frac{|\bm\kappa|}{\alpha} \left( e^{2\ii \alpha \gamma }-e^{2\ii\alpha}\right)\right]\nonumber
\end{align}
when the detectors' switching functions overlap \mbox{$T_B^\text{on}<T_A^\text{off}$}.

Inserting \eqref{S1int}, \eqref{S2nonint} and \eqref{S2overint} into \eqref{LS} and \eqref{MS}, \eqref{Lmunu} and \eqref{M} take the form
\begin{align}
\mathcal{L}_{AA}=&\frac{\lambda^2}{\pi^2}\int_0^\infty\text{d}|\bm{\kappa}|\,\frac{|\bm{\kappa}|e^{-\frac{1}{2}\bm{\kappa}^2\delta^2}}{\left(|\bm\kappa|+\alpha\right)^2}\sin^2\left[\frac{1}{2}(\alpha+|\bm\kappa|)\right],\\
\mathcal{L}_{AB}=-&\frac{\lambda^2}{4\pi^2\beta}\int_0^\infty\!\text{d}|\bm{\kappa}|\,\frac{e^{-\frac{1}{2}\bm{\kappa}^2\delta^2}}{\left(|\bm\kappa|+\alpha\right)^2}\sin(\beta|\bm{\kappa}|)\notag\\
&\times e^{-\ii(\gamma+1)(\alpha+|\bm\kappa|)}\left(e^{\ii(\alpha+|\bm\kappa|)}-1\right)^2,\\
|\mathcal{M}_\text{non}|=&\frac{\lambda^2}{4\pi^2\beta}\left|\int_0^\infty\text{d}|\bm{\kappa}|\,\frac{e^{-\frac{1}{2} \delta ^2 \bm{\kappa} ^2}e^{\ii\gamma(\alpha-|\bm\kappa|)}}{\bm{\kappa} ^2-\alpha ^2}\sin (\beta  |\bm{\kappa}| ) \right.\notag\\
&\times\left(e^{\ii (\alpha -\left| \bm{\kappa} \right| )}-1\right)\left(e^{\ii \left(\alpha +\left| \bm{\kappa} \right| \right)}-1\right)\Bigg|,\\
|\mathcal{M}_\text{over}|=&\frac{\lambda^2}{4\pi^2\beta}\Bigg|\int_0^{\infty}\text{d}|\bm{\kappa}|\,\sin (\beta  |\bm{\kappa}| )\,e^{-\frac{1}{2} \delta ^2 \bm{\kappa} ^2}{S_2}_\text{over}(\bm\kappa)\Bigg|,
\end{align}
where $\mathcal{M}_\text{over}$ and $\mathcal{M}_\text{non}$ correspond to the values of $\mathcal{M}$ when the detectors' switching functions overlap and when they do not, respectively.
\subsubsection{Sudden switching and quasi-pointlike detectors}

It is well known that a sudden switching for a pointlike detector in 3+1 dimensions leads to ultraviolet divergences in the response of a particle detector \cite{Louko2008}. However we can consider detectors whose spatial smearing is much smaller than the duration of the interaction (i.e. $\sigma/T\ll 1$) as an effective pointlike detector that can be switched abruptly in a 3+1 dimensional scenario.

\subsection{1+1 dimensions\label{1D}}
 As in the 3+1-dimensional case, we will explore the following different switching and spatial profile configurations for the detectors:
\begin{enumerate}
\item[B1.] Gaussian switching and Gaussian spatial smearing.
\item[B2.] Gaussian switching and pointlike detectors.
\item[B3.] Sudden switching and Gaussian spatial smearing.
\item[B4.] Sudden switching and pointlike detectors.
\end{enumerate}

\subsubsection{Gaussian switching functions and Gaussian smearing}
When we particularize \eqref{Lmunu} and \eqref{M} to 1+1 dimensions, the time integrals in \eqref{Lmunotint} and \eqref{Mnotint} are exactly the same as in the 3+1-dimensional scenario that we just computed. Therefore, in the 1+1-dimensional case, $\mathcal{L}_{\mu\nu}$ and $|\mathcal{M}|$ take the form
\begin{align}
\mathcal{L}_{AA}=&\frac{\tilde{\lambda}^2 e^{-\frac{1}{2}\alpha^2}}{4}\left[\int_{-\infty}^{-\Lambda}\text{d}\kappa\frac{1}{|{\kappa}|}e^{-\frac{1}{2}\kappa^2(1+\delta^2)}e^{-|\kappa|\alpha}\right.\label{LAA1DGS}\notag\\
&\left.+\int_\Lambda^\infty\text{d}\kappa\frac{1}{|{\kappa}|}e^{-\frac{1}{2}\kappa^2(1+\delta^2)}e^{-|\kappa|\alpha}\right],
\end{align}
\begin{align}
\mathcal{L}_{AB}=&\frac{\tilde{\lambda}^2 e^{-\frac{1}{2}\alpha^2} e^{-\ii\alpha\gamma}}{4}\notag\\
&\times\left[\int_{-\infty}^{-\Lambda}\text{d}\kappa\frac{e^{\ii\kappa\beta}}{|{\kappa}|}e^{-\ii|{\kappa}|\gamma}e^{-\frac{1}{2}\kappa^2(1+\delta^2)}e^{-|{\kappa}|\alpha}\right.\notag\\
&\left.+\int_\Lambda^\infty\text{d}\kappa\frac{e^{\ii\kappa\beta}}{|{\kappa}|}e^{-\ii|{\kappa}|\gamma}e^{-\frac{1}{2}\kappa^2(1+\delta^2)}e^{-|{\kappa}|\alpha}\right],\label{LAB1DGS}
\end{align}
\begin{align}
\left|\mathcal{M}_{\text{non}}\right|=&\frac{\tilde{\lambda}^2 e^{-\frac{1}{2}\alpha^2}}{4}\bigg|\int_{-\infty}^{-\Lambda}\text{d}\kappa\,\frac{e^{\ii\kappa\beta}}{|{\kappa}|}e^{\ii|{\kappa}|\gamma}e^{-\frac{1}{2}\kappa^2(1+\delta^2)}\notag\\
&+\int_\Lambda^\infty\text{d}\kappa\,\frac{e^{\ii\kappa\beta}}{|{\kappa}|}e^{\ii|{\kappa}|\gamma}e^{-\frac{1}{2}\kappa^2(1+\delta^2)}\bigg|,\label{Mnon1DGS}
\end{align}
\begin{align}
|\mathcal{M}_{\text{over}}|&=\frac{\tilde{\lambda}^2e^{-\frac{1}{2}\alpha^2}}{8}\notag\\
&\times\left|\int_{-\infty}^{-\Lambda}\!\!\!\!\!\!\!\!\text{d}\kappa\,\frac{e^{\ii\kappa\beta}}{|{\kappa}|}e^{-\frac{1}{2}\kappa^2\left(1+ \delta ^2\right)}\bigg[E(\kappa,\gamma)\!+\!E(\kappa,-\gamma)\bigg]\right.\notag\\
&+\!\!\!\int_\Lambda^\infty\!\!\!\!\!\!\text{d}\kappa\,\frac{e^{\ii\kappa\beta}}{|{\kappa}|}e^{-\frac{1}{2}\kappa^2 (1+\delta ^2)}\bigg[E(\kappa,\gamma)+E(\kappa,-\gamma)\bigg]\!\Bigg|\label{Mover1DGS},
\end{align}
where the notation $\mathcal{M}_\text{over}$ and $\mathcal{M}_\text{non}$ is understood in the same way as in section \ref{3D}, and $\Lambda$ is an infrared cutoff which regularizes the well-known logarithmic IR divergences of the 1+1-dimensional case. This kind of IR regularization is common in the literature and can be thought, for instance, as the length scale of a very long optical cavity or a periodic optical fiber, or the characteristic radius of a cylinder spacetime topology.

Notice that, as opposed to the 3+1-dimensional case (where the field has units of $[\phi]=T^{-1}$ and $\lambda$ is dimensionless), in 1+1 dimensions the field is unitless, hence $\lambda$ has units of $T$. Therefore, in Eqs. \eqref{LAA1DGS}, \eqref{LAB1DGS}, \eqref{Mnon1DGS} and \eqref{Mover1DGS} we have already used a dimensionless coupling strength defined as $\tilde{\lambda}=\lambda T$.
\subsubsection{Gaussian switching functions and pointlike detectors}
As in the 3+1-dimensional scenario, the pointlike case corresponds to the limit $\delta\rightarrow 0$ in Eqs. \eqref{LAA1DGS}, \eqref{LAB1DGS}, \eqref{Mnon1DGS} and \eqref{Mover1DGS}, yielding
\begin{align}
\mathcal{L}_{AA}=&\frac{\tilde{\lambda}^2 }{4}\!\!\left[\int_{-\infty}^{-\Lambda}\!\!\!\!\!\!\!\!\text{d}\kappa\frac{e^{-\frac{1}{2}\left(\alpha+|\kappa|\right)^2}}{|\kappa|}\!+\!\!\!\int_\Lambda^\infty\!\!\!\!\!\text{d}\kappa\frac{e^{-\frac{1}{2}\left(\alpha+|\kappa|\right)^2}}{|{\kappa}|}\right]\!\!,
\end{align}
\begin{align}
\mathcal{L}_{AB}=&\frac{\tilde{\lambda}^2 e^{-\frac{1}{2}\alpha^2} e^{-\ii\alpha\gamma}}{4}\left[\int_{-\infty}^{-\Lambda}\text{d}\kappa\frac{e^{\ii\kappa\beta}}{|{\kappa}|}e^{-\ii|{\kappa}|\gamma}e^{-\frac{1}{2}\kappa^2}e^{-|{\kappa}|\alpha}\right.\notag\\
&\left.+\int_\Lambda^\infty\text{d}\kappa\frac{e^{\ii\kappa\beta}}{|{\kappa}|}e^{-\ii|{\kappa}|\gamma}e^{-\frac{1}{2}\kappa^2}e^{-|{\kappa}|\alpha}\right],\\
\left|\mathcal{M}_{\text{non}}\right|=&\frac{\tilde{\lambda}^2 e^{-\frac{1}{2}\alpha^2}}{4}\bigg|\int_{-\infty}^{-\Lambda}\text{d}\kappa\,\frac{e^{\ii\kappa\beta}}{|{\kappa}|}e^{\ii|{\kappa}|\gamma}e^{-\frac{1}{2}\kappa^2}\notag\\
&+\int_\Lambda^\infty\text{d}\kappa\,\frac{e^{\ii\kappa\beta}}{|{\kappa}|}e^{\ii|{\kappa}|\gamma}e^{-\frac{1}{2}\kappa^2}\bigg|,\\
|\mathcal{M}_{\text{over}}|&=\frac{\tilde{\lambda}^2e^{-\frac{1}{2}\alpha^2}}{8}\notag\\
&\left|\int_{-\infty}^{-\Lambda}\!\!\!\!\!\!\!\!\text{d}\kappa\,\frac{e^{\ii\kappa\beta}}{|{\kappa}|}e^{-\frac{1}{2}\kappa^2}\bigg[E(\kappa,\gamma)+E(\kappa,-\gamma)\bigg]\right.\notag\\
&+\!\!\!\int_\Lambda^\infty\!\!\!\!\!\!\text{d}\kappa\,\frac{e^{\ii\kappa\beta}}{|{\kappa}|}e^{-\frac{1}{2}\kappa^2}\bigg[E(\kappa,\gamma)+E(\kappa,-\gamma)\bigg]\!\Bigg|.
\end{align}

\subsubsection{Sudden switching functions and Gaussian smearing}
In this case the time integrals are again \eqref{S1} and \eqref{S2}. Inserting them into \eqref{Lmunotint} and \eqref{Mnotint} along with the smearing function \eqref{smear}, \eqref{Lmunu} and \eqref{M} now read:
\begin{align}
\mathcal{L}_{AA}=&\frac{\tilde{\lambda}^2}{\pi}\left\{\int_{-\infty}^{-\Lambda}\text{d}\kappa\, \frac{e^{-\frac{1}{2}\kappa^2\delta^2}}{|\kappa|\left(|\kappa|+\alpha\right)^2}\sin^2\left[\frac{1}{2}(\alpha+|\kappa|)\right]\right.\notag\\
&\left.+\int_\Lambda^\infty\text{d}\kappa\, \frac{e^{-\frac{1}{2}\kappa^2\delta^2}}{|\kappa|\left(|\kappa|+\alpha\right)^2}\sin^2\left[\frac{1}{2}(\alpha+|\kappa|)\right]\right\}\label{LAA1DSS},
\end{align}
\begin{align}
\mathcal{L}_{AB}=&-\frac{\tilde{\lambda}^2}{4\pi}\left[\int_{-\infty}^{-\Lambda}\text{d}\kappa\, \frac{e^{-\frac{1}{2}\kappa^2\delta^2}e^{\ii\kappa\beta}}{|\kappa|\left(|\kappa|+\alpha\right)^2}\right.\notag\\
&\times  e^{-\ii (\gamma +1) (\alpha +\left| \kappa \right| )}\left(e^{\ii (\alpha +\left| \kappa \right| )}-1\right)^2\notag\\
&+\int_\Lambda^\infty\text{d}\kappa\, \frac{e^{-\frac{1}{2}\kappa^2\delta^2}e^{\ii\kappa\beta}}{|\kappa|\left(|\kappa|+\alpha\right)^2}\notag\\
&\times  e^{-\ii (\gamma +1) (\alpha +\left| \kappa \right| )}\left(e^{\ii (\alpha +\left| \kappa \right| )}-1\right)^2\Bigg]\label{LAB1DSS},
\end{align}
\begin{align}
|\mathcal{M}_\text{non}|=&\frac{\tilde{\lambda}^2}{4\pi}\left|\int_{-\infty}^{-\Lambda}\text{d}\kappa\frac{e^{\ii \kappa \beta}e^{\ii\gamma(\alpha-|\kappa|)}}{|\kappa|\left(\left| \kappa \right| ^2-\alpha ^2\right)}e^{-\frac{1}{2}\kappa^2\delta^2}\right.\notag\\
&\times\left(e^{\ii (\alpha -\left| \kappa \right| )}-1\right)\left(e^{\ii \left(\alpha +\left| \kappa \right| \right)}-1\right)\notag\\
&+\int_{\Lambda}^{\infty}\text{d}\kappa\frac{e^{\ii \kappa \beta}e^{\ii\gamma(\alpha-|\kappa|)}}{|\kappa|\left(\left| \kappa \right| ^2-\alpha ^2\right)}e^{-\frac{1}{2}\kappa^2\delta^2}\notag\\
&\left(e^{\ii (\alpha -\left| \kappa \right| )}-1\right)\left(e^{\ii \left(\alpha +\left| \kappa \right| \right)}-1\right)\!\Bigg|,\label{Mnon1DSS}
\end{align}
\begin{align}
|\mathcal{M}_\text{over}|=&\frac{\tilde{\lambda}^2}{4\pi}\Bigg|\int_{-\infty}^{-\Lambda}\text{d}\kappa\,\frac{e^{\ii\kappa\beta}}{|\kappa|}e^{-\frac{1}{2} \delta ^2\kappa ^2}{S_2}_{over}(\kappa)\notag\\
&+\int_{\Lambda}^{\infty}\text{d}\kappa\,\frac{e^{\ii\kappa\beta}}{|\kappa|}e^{-\frac{1}{2} \delta ^2 \kappa ^2}{S_2}_{over}(\kappa)\Bigg|\label{Mover1DSS}.
\end{align}

\subsubsection{Sudden switching and pointlike detectors}

In 1+1 dimensions the pointlike limit of the previous case is not divergent, and we can therefore safely take the limit $\delta\rightarrow 0$ in Eqs. \eqref{LAA1DSS}, \eqref{LAB1DSS}, \eqref{Mnon1DSS} and \eqref{Mover1DSS}, which results in
\begin{align}
\mathcal{L}_{AA}=&\frac{\tilde{\lambda}^2}{\pi}\left\{\int_{-\infty}^{-\Lambda}\text{d}\kappa\, \frac{1}{|\kappa|\left(|\kappa|+\alpha\right)^2}\sin^2\left[\frac{1}{2}(\alpha+|\kappa|)\right]\right.\notag\\
&+\left.\int_\Lambda^\infty\text{d}\kappa\, \frac{1}{|\kappa|\left(|\kappa|+\alpha\right)^2}\sin^2\left[\frac{1}{2}(\alpha+|\kappa|)\right]\right\},
\end{align}
\begin{align}
\mathcal{L}_{AB}=&-\frac{\tilde{\lambda}^2}{4\pi}\left[\int_{-\infty}^{-\Lambda}\!\!\!\!\!\!\!\text{d}\kappa\, \frac{e^{\ii\kappa\beta} e^{-\ii (\gamma +1) (\alpha +\left| \kappa \right| )}}{|\kappa|\left(|\kappa|+\alpha\right)^2}\!\left(e^{\ii (\alpha +\left| \kappa \right| )}\!-\!1\!\right)^2\right.\notag\\
&+\left.\int_\Lambda^\infty\!\!\text{d}\kappa\, \frac{e^{\ii\kappa\beta} e^{-\ii (\gamma +1) (\alpha +\left| \kappa \right| )}}{|\kappa|\left(|\kappa|+\alpha\right)^2}\left(e^{\ii (\alpha +\left| \kappa \right| )}-1\right)^2\right],
\end{align}
\begin{align}
|\mathcal{M}_\text{non}|=&\frac{\tilde{\lambda}^2}{4\pi}\left|\int_{-\infty}^{-\Lambda}\text{d}\kappa\frac{e^{\ii \kappa \beta}e^{\ii\gamma(\alpha-|\kappa|)}}{|\kappa|\left(\left| \kappa \right| ^2-\alpha ^2\right)}\right.\notag\\
&\times\left(e^{\ii (\alpha -\left| \kappa \right| )}-1\right)\left(e^{\ii \left(\alpha +\left| \kappa \right| \right)}-1\right)\notag\\
&+\int_{\Lambda}^{\infty}\text{d}\kappa\frac{e^{\ii \kappa \beta}e^{\ii\gamma(\alpha-|\kappa|)}}{|\kappa|\left(\left| \kappa \right| ^2-\alpha ^2\right)}\notag\\
&\times\left(e^{\ii (\alpha -\left| \kappa \right| )}-1\right)\left(e^{\ii \left(\alpha +\left| \kappa \right| \right)}-1\right)\!\Bigg|,
\end{align}
\begin{align}
|\mathcal{M}_\text{over}|\!=&\frac{\tilde{\lambda}^2}{4\pi}\Bigg|\!\int_{-\infty}^{-\Lambda}\!\!\!\!\!\!\!\text{d}\kappa\,\frac{e^{\ii\kappa\beta}}{|\kappa|}{S_2}_{over}(\kappa)\!+\!\!\int_{\Lambda}^{\infty}\!\!\!\!\!\text{d}\kappa\,\frac{e^{\ii\kappa\beta}}{|\kappa|}{S_2}_{over}(\kappa)\Bigg|.
\end{align}

We have now all the ingredients needed to study both entanglement and general correlations from a quantum scalar field in 3+1 and 1+1 flat spacetimes for the different switching functions and spatial smearings considered.

\section{Entanglement harvesting}\label{entharvest}

We are interested in quantifying the entanglement acquired by particle detectors after their interaction with the vacuum state of the field. As we will see, factors such as the dimensionality of spacetime, the nature of the switching functions, the smearing of the detectors and their internal energy structure have a dramatic impact on their ability to harvest field vacuum entanglement.

We will use negativity \cite{Vidal2002} to quantify entanglement. For a two-qubit system, the negativity (defined as the sum of the negative eigenvalues of the partially transposed density matrix) is an entanglement monotone which only vanishes for separable states \cite{Horodecki1996,Peres1996}.

\subsection{Entanglement harvesting to second order in perturbation theory} \label{sdf}

To second order in perturbation theory there is only one eigenvalue of the partial transpose of \eqref{eq:seconddens} that can be negative
\begin{equation}
E_1\!=\frac{1}{2}\left[\mathcal{L}_{AA}\!+\!\mathcal{L}_{BB}\!-\!\sqrt{\left(\mathcal{L}_{AA}\!-\!\mathcal{L}_{BB}\right)^2\!+\!4\left|\mathcal{M}\right|^2}\right]\!+\!\mathcal{O}(\lambda_\nu^4).
\end{equation}

Note that a naive inspection of the partial transpose of \eqref{eq:seconddens} would have produced the apparently always negative eigenvalue $E_2\!=-\left|\mathcal{L}_{AB}\right|^2$. However note that $\left|\mathcal{L}_{AB}\right|^2$ is $\mathcal{O}\left(\lambda_\nu^4\right)$, and thus $E_2=0+\mathcal{O}\left(\lambda_\nu^4\right)$. As we will discuss later, to find the correct form of $E_2$ the whole fourth order correction to the density matrix must be computed, resulting in $E_2$ not being negative in general. Therefore, we define the following negativity estimator:
\begin{align}
\mathcal{N}^{(2)}=&-E_1\notag\\
=&-\frac{1}{2}\left[\mathcal{L}_{AA}+\mathcal{L}_{BB}-\sqrt{\left(\mathcal{L}_{AA}-\mathcal{L}_{BB}\right)^2+4\left|\mathcal{M}\right|^2}\right],
\end{align}
so that the negativity, to second order in the coupling strength, is equal to $\mathcal{N}=\text{max}(0,\,\mathcal{N}^{(2)})$.

This form of the negativity estimator can be further simplified when it is assumed that the two detectors are identical and switched on for the same amount of time (but with a time delay between them), and therefore $\mathcal{L}_{\mu\mu}\equiv\mathcal{L}_{AA}=\mathcal{L}_{BB}$. Under this assumption, at leading order in perturbation theory the previous formula can be written as
\begin{equation}
\mathcal{N}^{(2)}=|\mathcal{M}|-\mathcal{L}_{\mu\mu}. \label{negat2equal}
\end{equation}

This expression shows very intuitively that for the state to be entangled, the nonlocal term $\mathcal{M}$ has to be larger than the local terms $\mathcal{L}_{\mu\mu}$ \cite{Reznik2005}. We can now evaluate the impact of the smoothness of the switching on the ability of the detectors to harvest entanglement from the vacuum. In view of \eqref{negat2equal}, it is not straightforward what the impact of the switching may be: a sudden switching may, on the one hand, increase the local noise (the particle count of the individual detectors, $\mathcal{L}_{AA}$ and $\mathcal{L}_{BB}$), but on the other hand it is not \textit{a priori} clear what would be the impact of the switching on the nonlocal terms $\mathcal{M}$. Therefore we will need to compare the two scenarios (smooth vs. sudden switching) to confirm if the extra noise introduced by a sudden switching hinders the detectors' ability to harvest entanglement.

\begin{figure*}
\begin{tabular}{cccc}
\includegraphics[width=0.21\textwidth]{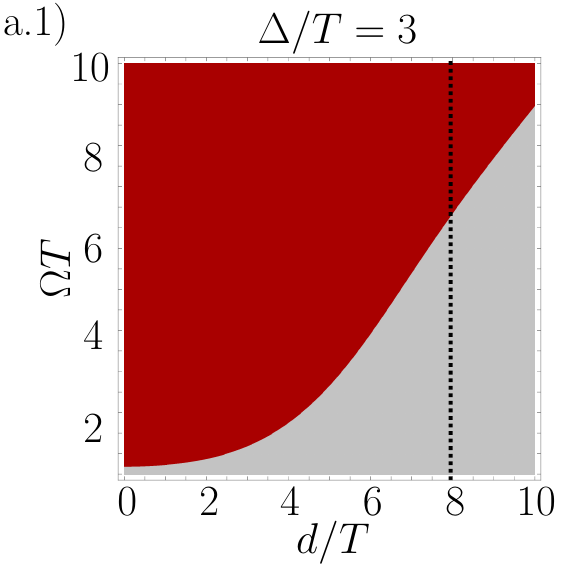}
&\includegraphics[width=0.27\textwidth]{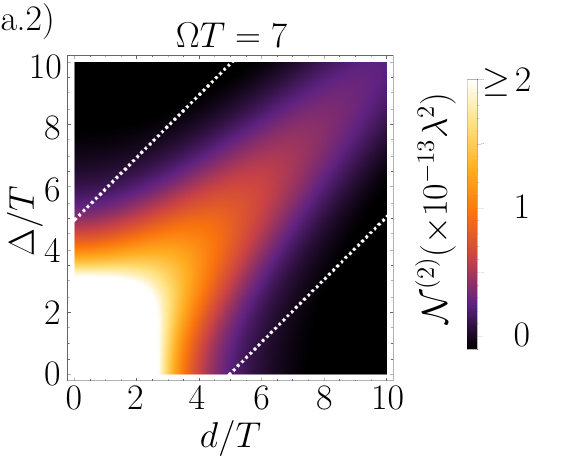}
&\includegraphics[width=0.21\textwidth]{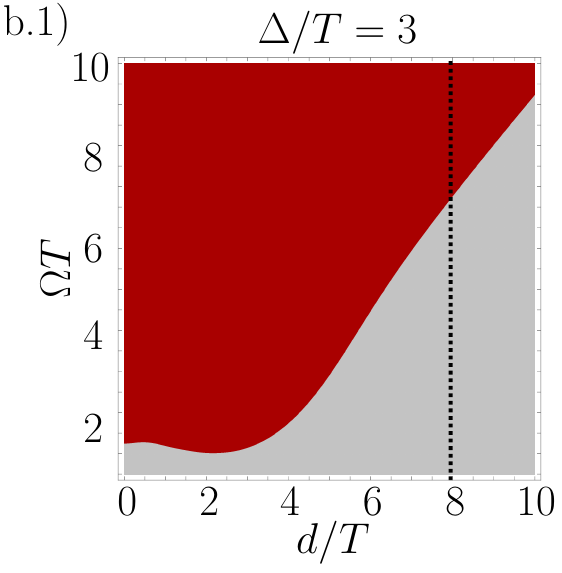}
&\includegraphics[width=0.27\textwidth]{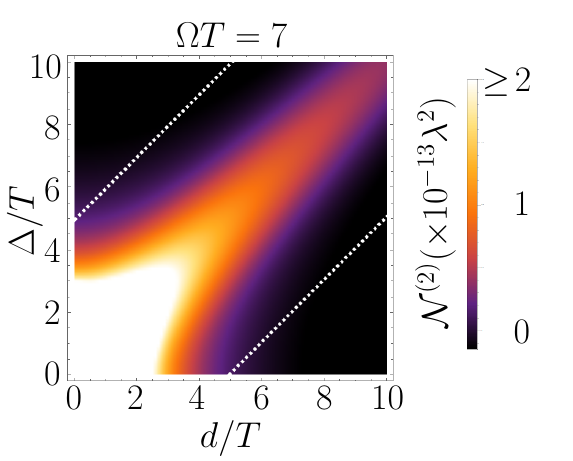}\\
\includegraphics[width=0.21\textwidth]{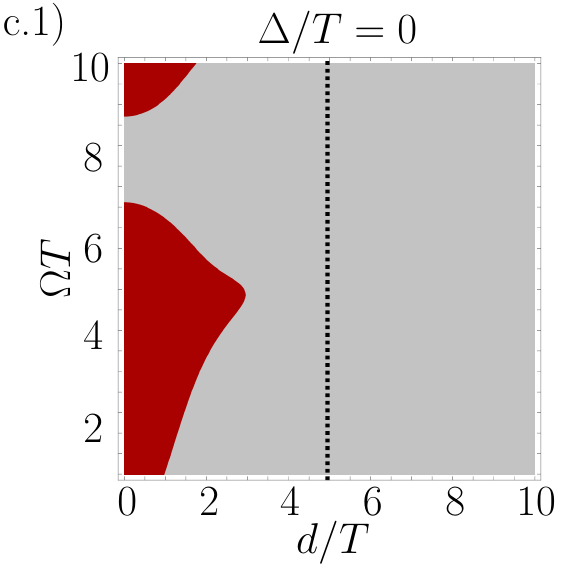}
&\includegraphics[width=0.27\textwidth]{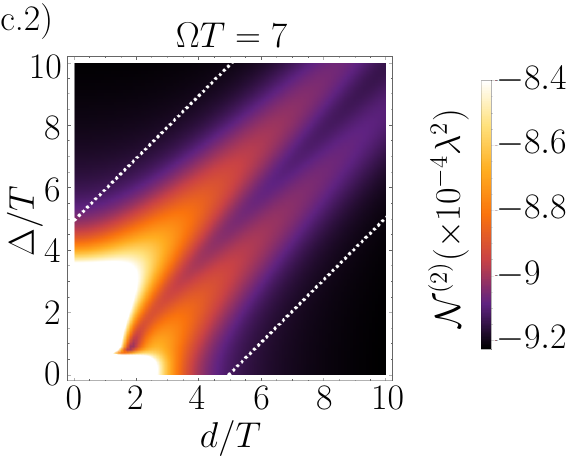}
& \includegraphics[width=0.21\textwidth]{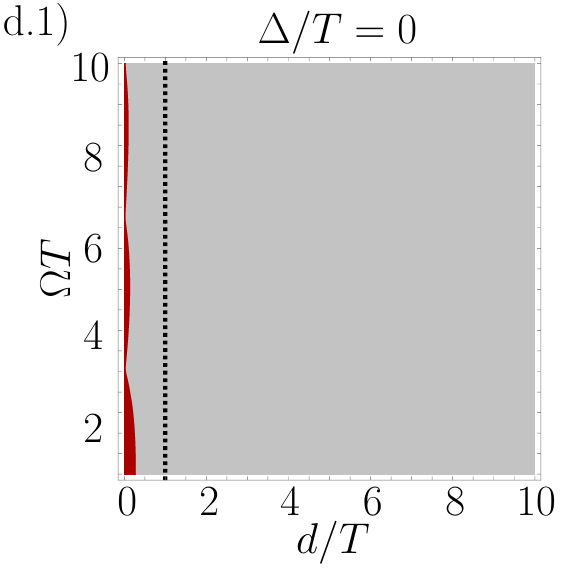}
& \includegraphics[width=0.28\textwidth]{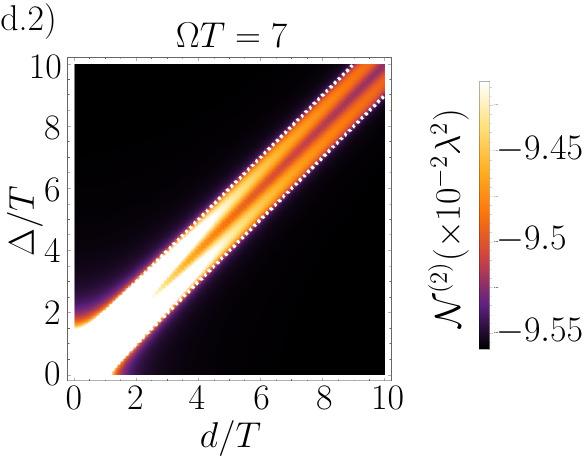}\\
\includegraphics[width=0.21\textwidth]{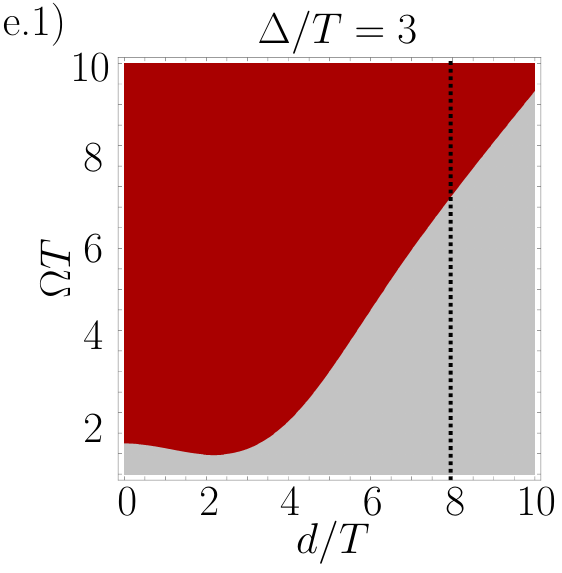}
& \includegraphics[width=0.27\textwidth]{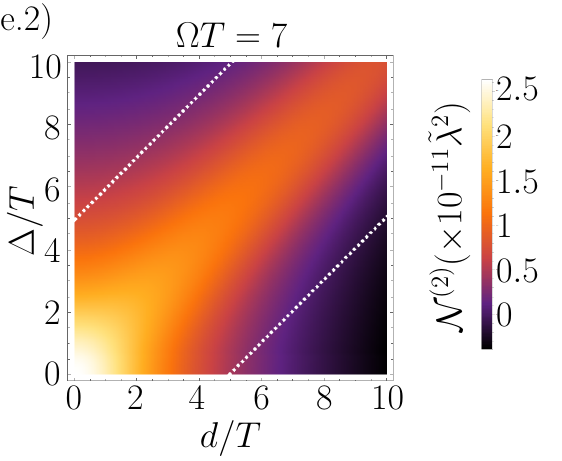}
&\includegraphics[width=0.21\textwidth]{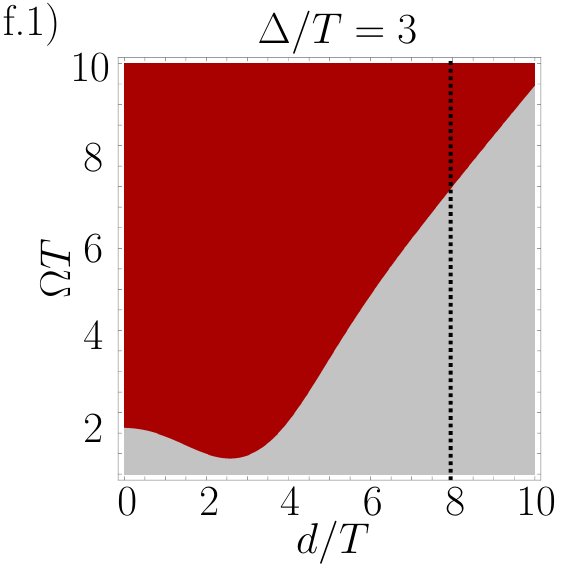}
&\includegraphics[width=0.27\textwidth]{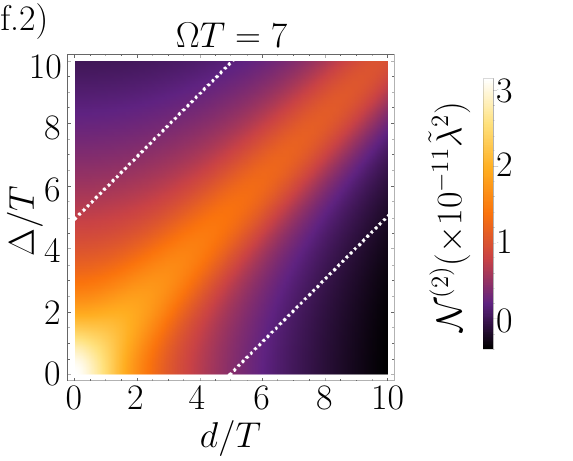}\\
\includegraphics[width=0.21\textwidth]{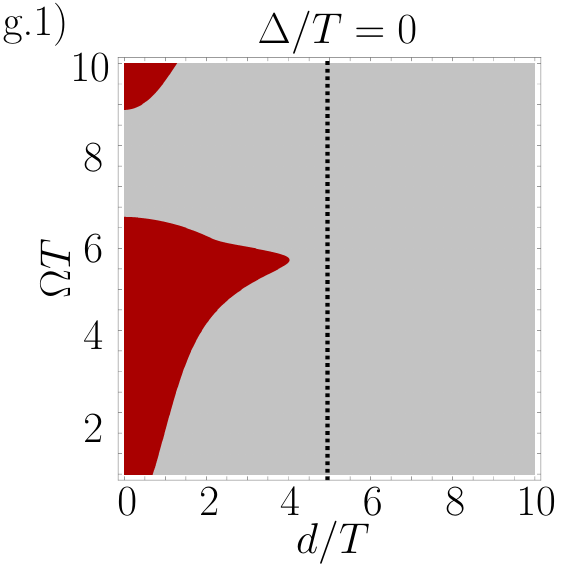}
&\includegraphics[width=0.27\textwidth]{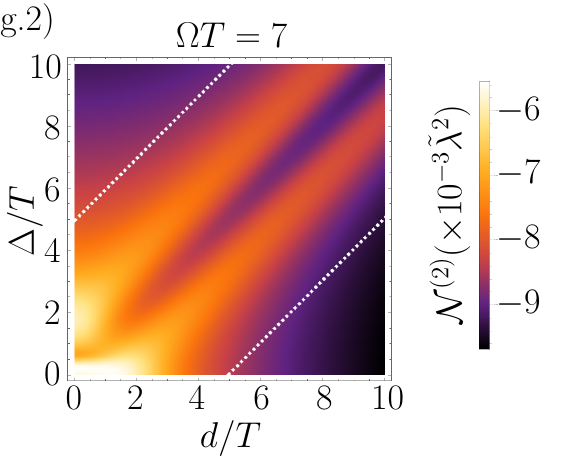}
&\includegraphics[width=0.21\textwidth]{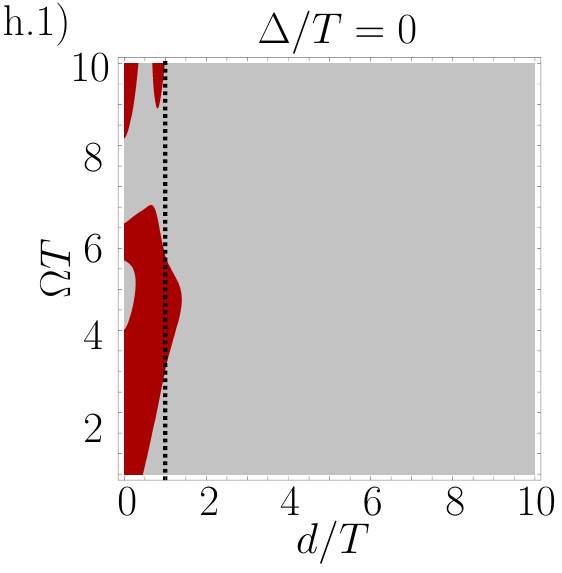}
&\includegraphics[width=0.27\textwidth]{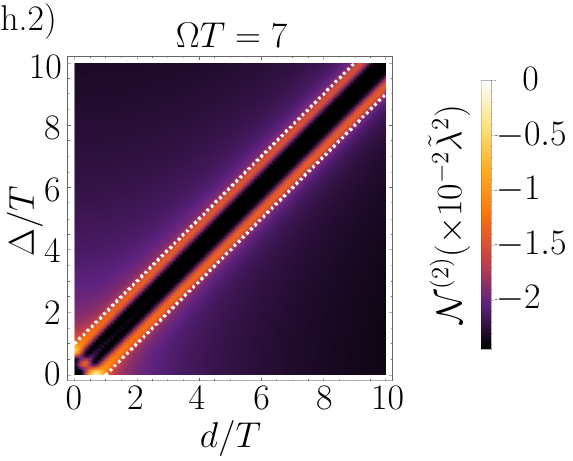}\\
\end{tabular}
\caption{Negativity for all the cases described in sections \ref{3D} and \ref{1D}, in the following order: (a) three-dimensional (3D) Gaussian switching and Gaussian spatial profile; (b) 3D Gaussian switching and pointlike detectors; (c) 3D sudden switching and Gaussian spatial profile; (d) 3D sudden switching and near-pointlike detectors ($\sigma/T\ll1$); (e) one-dimensional (1D) Gaussian switching and Gaussian spatial profile; (f) 1D Gaussian switching and pointlike detectors; (g) 1D sudden switching and Gaussian spatial profile; (h) 1D sudden switching and pointlike detectors. The plots a.1), b.1), $\dots$, h.1) (first and third columns) show in (dark) red the values of $d/T$ and $\Omega T$ for which entanglement harvesting is possible, and in (light) grey the values for which there is no entanglement harvesting. The plots denoted by a.2), b.2), $\dots$, h.2) (second and fourth columns) show the specific value of the negativity estimator $\mathcal{N}^{(2)}$ as a function of the spatial separation of the detectors and the time delay between their switching functions for a given value of $\Omega T$ (recall that there is entanglement harvesting when $\mathcal{N}^{(2)}>0$). In all plots the dashed lines represent the boundaries of the lightcone. In the second and fourth columns, the points to the right of the rightmost dark line represent spacelike separation. These boundaries are placed at $\Delta=d\pm T$ in the cases of pointlike detectors which are switched suddenly, and at $\Delta=d\pm 7T/\sqrt{2}$ in the rest, which is a reasonable estimation given the discussion in section II, subsection A1. All cases of Gaussian spatial profile have $\sigma/T=1$ except for the near-pointlike detectors (plots d.1 and d.2), for which $\sigma/T=0.01$. We see that Gaussian switching always allows for spacelike entanglement harvesting for sufficiently large values of $\Omega T$, while this is not the case for sudden switching. For the 1+1 dimensional scenarios the IR cutoff was taken $\Lambda T=0.001$, much smaller than all the relevant frequency scales in the setup.}
\label{negats}
\end{figure*}

In Fig. \ref{negats} we present some examples of the negativity in all the cases defined in sections \ref{3D} and \ref{1D}. For the binary plots showing the parameter region where entanglement harvesting is possible (a.1, b.1, c.1, d.1, e.1, f.1, g.1, h.1), we have chosen to fix $\Delta/T=0$ (no delay between the switchings) for the sudden switching cases because this maximizes the parameter region where entanglement harvesting is possible. This is because abruptly switched detectors quickly lose their ability to extract vacuum entanglement as  $\Delta/T$ grows, even in lightlike connection. For the Gaussian switching this is not the case, and therefore we have chosen to show the parameter regions of entanglement harvesting for $\Delta/T=3$ in order to better display the features of the parameter dependence.

The plots (a.2, b.2, c.2, d.2, e.2, f.2, g.2, h.2) show the variation in the negativity estimator $\mathcal{N}^{(2)}$ with the distance between the detectors ($d/T$) and their switching delay ($\Delta/T$) for a representative value of \mbox{$\Omega T=7$}. Notice that, since the negativity is defined as \mbox{$\mathcal{N}=\text{max}(\mathcal{N}^{(2)},0)+\mathcal{O}(\lambda_\nu^4) $}, in the cases where the plots display negative values there is no entanglement harvesting at all. It is still illustrative in those cases (all of them sudden switching) to show how far from zero the negativity estimator $\mathcal{N}^{(2)}$ would be.

First, we note that the entanglement peaks in the region where there is lightlike contact between the detectors (in between of the two white, dashed lines). In this situation both detectors can exchange real quanta via the background field, resulting in a peak in the negativity that will be also visible in the study of the total correlations. In the majority of the cases negativity peaks when there is full lightlike contact (exactly at the exact center of the region inside the two white lines in Fig. \ref{negats}). However, notice that there are two different trends: the entanglement increases when the detectors are lightlike connected, but decreases in a nontrivial way with the spatiotemporal distance, thus it is not surprising that in some cases (for instance, for the 1+1-dimensional cases with sudden switching functions) the maxima of the negativity estimator do not appear at the center of the lightlike region, although these maxima are still always somewhere in the region of light contact.

Another clear feature that arises from our analysis is that, perhaps contrary to intuition, the detectors switched on in a smooth, more adiabatic manner are capable of harvesting entanglement in a much more efficient way than suddenly switched detectors, both when the detectors are in causal contact and when they are spacelike separated.

Let us focus now on the interesting case of spacelike entanglement harvesting. Since the detectors cannot exchange information, spacelike entanglement is the result of the harvesting of only preexisting correlations in the quantum vacuum.

We see from Fig. \ref{negats} that for smooth Gaussian switching it is always possible to obtain spacelike entanglement harvesting if the internal energy gap of the detectors $\Omega$ is increased (consistent with the analysis in \cite{Salton2015}), but paying the price that the larger the value of $\Omega$, the smaller (but finite) the amount of entanglement harvested. In other words, for Gaussianly switched detectors, when the energy gap increases the total amount of entanglement decreases rapidly, but in exchange the region of nonzero negativity increases, eventually ``leaking'' arbitrarily much into the spacelike-separation region.

Strikingly, this is not the case for sudden switching, where the behaviour of negativity with the internal gap of the detectors is dramatically different to the Gaussian switching case.

In particular, for sudden switching, negativity is no longer monotonic with the detectors' energy gap and, what is more, the values of $\Omega$ that allow for spacelike entanglement harvesting are severely limited. We found that there are even cases in which it is impossible to harvest entanglement with spacelike-separated suddenly-switched detectors for any value of $\Omega$ within the range explored. Increasing values of $\Omega$ do not seem to improve this situation (see Figs. \ref{negats}c, \ref{negats}d, \ref{negats}g); it is only possible for pointlike detectors in 1+1 dimensions to find spacelike entanglement harvesting with a sudden switching, and increasing $\Omega$ actually reduces the spacelike harvesting ability even in this case (see Fig. \ref{negats}h).

\begin{figure*}[t]
\centering
\begin{tabular}{ccc}
\includegraphics[width=0.32\textwidth]{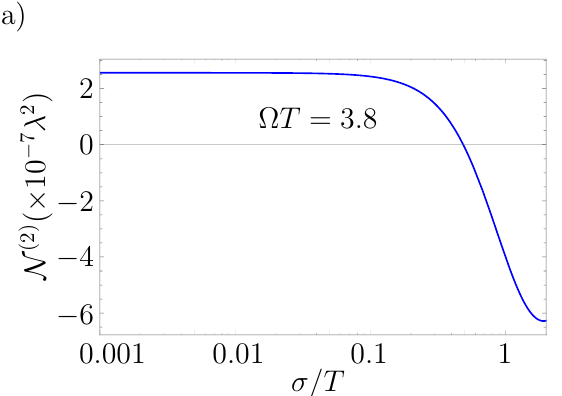}
&\includegraphics[width=0.32\textwidth]{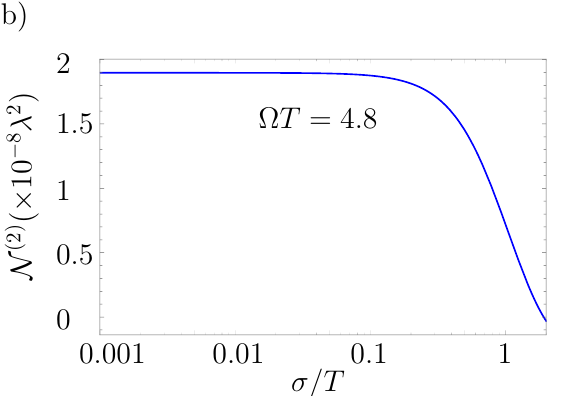}
&\includegraphics[width=0.32\textwidth]{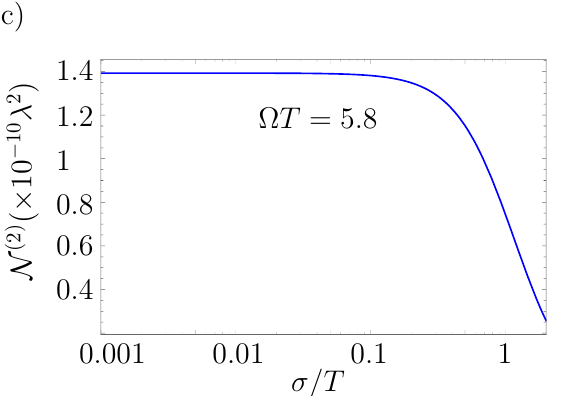}
\end{tabular}
\caption{Dependence of entanglement harvesting on the size of the detectors for various values of the internal energy gap. The detectors are located in a 3+1-dimensional flat spacetime, and are switched on using the Gaussian profile \eqref{gauss}. We see that the larger the detectors, the less efficient they are for entanglement harvesting.}
\label{detecsize}
\end{figure*}

The conclusion we extract is that the increase of the local noise as a result of a sudden switching hinders our ability to harvest quantum entanglement from the field, regardless of any possible beneficial effect that the sudden switching may have in the nonlocal $\mathcal{M}$, whether the detectors are or not in causal contact or the dimension of spacetime.

Finally, let us consider how our ability to harvest entanglement from the vacuum varies with the size of the detectors. This behaviour is shown in Fig. \ref{detecsize}. First note how the behaviour of the detector quickly goes to the pointlike limit when $\sigma/T\ll 1$.

For larger $\sigma$ we see that as the sizes of the detectors grow, they become less efficient to harvest vacuum entanglement. This is true as long as $\sigma\lesssim 6 d$, this is, when the two detector Gaussian smearings are further apart than about 8 standard deviations. Otherwise, if we were to allow the detectors to have a spatial overlap (a rather unphysical situation, however), this overlap enhances the ability of the detectors to get entangled.

\subsection{Entanglement harvesting beyond second order in perturbation theory}

Recall that in the analysis of the negativity that we carried out in section \ref{sdf} we found that there was an eigenvalue of the partially transposed density matrix, $E_2$, which was $\mathcal{O}(\lambda_\nu^4)$. This eigenvalue `naively' appeared to be always negative when taking into account only the contributions coming from the square of second-order perturbative terms. However, to find the right fourth-order perturbative corrections to the negativity (and in particular, the right value of $E_2$) it is not enough to cut the perturbative expansion at second order as in \eqref{eq:seconddens}. If we expand $\rho_{AB}$ to $\mathcal{O}(\lambda_\nu^4)$ we obtain
\begin{equation}
\rho_{AB}=\begin{pmatrix}
\rho^{\textsc{iv}}_{11} & 0 & 0 & \rho^{\textsc{iv}}_{14} \\
0 & \rho^{\textsc{iv}}_{22} &\rho^{\textsc{iv}}_{23}& 0 \\
0 &  {\rho^{\textsc{iv}}_{23}}^*  & \rho^{\textsc{iv}}_{33} & 0 \\
{\rho^{\textsc{iv}}_{14}}^* & 0 & 0 & \rho^{\textsc{iv}}_{44}
\end{pmatrix}+\mathcal{O}(\lambda_\nu^6),\label{fourthdens}
\end{equation}
where
\begin{align}
\nonumber\rho^{\textsc{iv}}_{11}=&1-\left(\mathcal{L}_{AA}+\mathcal{L}_{BB}\right)-\left(\Xi_1+\Xi_2+\Xi_3\right),\\
\nonumber\rho^{\textsc{iv}}_{22}=& \mathcal{L}_{AA}+\Xi_1,\\
\rho^{\textsc{iv}}_{33}=&\mathcal{L}_{BB}+\Xi_2,\\
\nonumber\rho^{\textsc{iv}}_{44}=&\Xi_3,\\
\nonumber\rho^{\textsc{iv}}_{14}=&\mathcal{M}^*+\Upsilon,\\
\nonumber\rho^\textsc{iv}_{23}=&\mathcal{L}_{AB} +\Pi.
\end{align}

Here, all the Greek letters denote terms which are proportional to the fourth power of the coupling strength. The partial transpose of \eqref{fourthdens} has two potentially negative eigenvalues:
\begin{align}
E_1=&\frac{1}{2}\Bigg[\left(\mathcal{L}_{AA}+\mathcal{L}_{BB}-\sqrt{\left(\mathcal{L}_{AA}-\mathcal{L}_{BB}\right)^2+4 \left|\mathcal{M}\right|^2}\right)\notag\\
+&\Bigg(\Xi_1+\Xi_2-\frac{4\text{Re}(\Upsilon\mathcal{M})}{\sqrt{4\left|\mathcal{M}\right|^2+\left(\mathcal{L}_{AA}-\mathcal{L}_{BB}\right)^2}}\notag\\
-&\frac{\left(\mathcal{L}_{AA}-\mathcal{L}_{BB}\right) \left(\Xi_1-\Xi_2\right)}{\sqrt{4\left|\mathcal{M}\right|^2+\left(\mathcal{L}_{AA}-\mathcal{L}_{BB}\right)^2}}\Bigg)\Bigg]\!+\!\mathcal{O}(\lambda_\nu^6),\\
E_2=&\left(\Xi _3-|\mathcal{L}_{AB}|^2\right)+\mathcal{O}(\lambda_\nu^6). \label{E2}
\end{align}

$E_1$ is just the eigenvalue found in the study of the second-order correction with additional fourth-order terms. These fourth-order corrections are generally smaller than the second-order ones, so they will only play a role when the second-order contributions are zero.

We also find that $E_2$ is actually not always negative, and its sign will depend on the value of two competing terms, one of which is expressed in terms of \eqref{Lmunu} and the other one being
\begin{align}
\Xi_3\!=\!\!\int\!\!\text{d}^n\bm{k}\!\!\int\!\!\text{d}^n\bm{q}\left[\xi_1(\bm k,\bm q)\!+\!\xi_2(\bm k,\bm q)\!+\!\xi_3(\bm k,\bm q)\!+\!\xi_4(\bm k,\bm q)\right],
\end{align}
where each of the $\xi_i(\bm\kappa)$ are given in Appendix \ref{computeXi}.

\begin{figure}[h]
\centering
\begin{tabular}{c}
\includegraphics[width=0.45\textwidth]{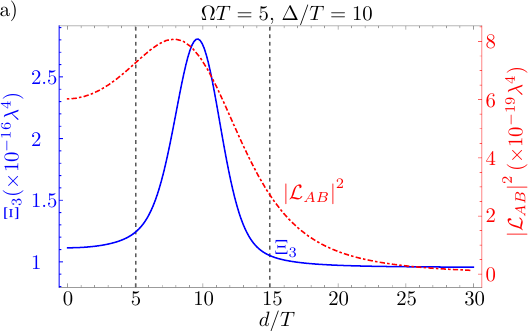}\\
\includegraphics[width=0.45\textwidth]{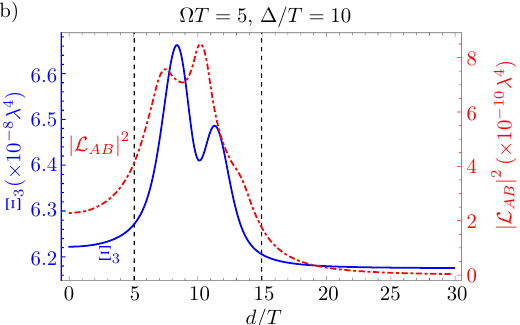}
\end{tabular}
\caption{$\Xi_3$ and $\left|\mathcal{L}_{AB}\right|^2$ for the cases of spatially smeared detectors in 3+1 dimensions and a) Gaussian and b) sudden switching functions, for a time delay of $\Delta=10T$ between their switchings. The vertical dashed lines represent the boundaries of the lightcone. Recall that there is a fourth-order contribution to entanglement harvesting only if $|\mathcal{L}_{AB}|^2>\Xi_3$, which never happens in these cases. Notice the use of different $y$-axis scales for $|\mathcal{L}_{AB}|^2$ and $\Xi_3$.}
\label{fourth}
\end{figure}

It can be checked that, as a general trend, $E_2$ does not become negative when $E_1$ is positive (i.e, when \mbox{$\mathcal{N}^{(2)}\leq0$}), therefore the fourth-order analysis of entanglement harvesting does not add significatively new results. For illustration, we show in Fig. \ref{fourth} the two competing terms $\Xi_3$ and $\left|\mathcal{L}_{AB}\right|^2$ in \eqref{E2} for the case of 3+1 dimensions with sudden and Gaussian switching. Recall that there would only be fourth-order contribution to entanglement if $|\mathcal{L}_{AB}|^2>\Xi_3$.

\section{Harvesting of mutual information}\label{corrharvest}

In the previous section we have shown that entanglement harvested between two two-level particle detectors strongly depends on the detectors' switching, their spatial smearing, their internal energy gap and the dimensionality of spacetime.

It is also an interesting question to find how particle detectors can harvest general correlations from the field, and not only entanglement. For instance, is it possible to harvest classical correlations from the field while the detectors remain spacelike separated?

In this section we will show that the harvesting of correlations from the field is much easier than harvesting entanglement. Namely, a pair of detectors prepared in their ground state can harvest correlations for a much wider range of scenarios than those that allow for entanglement harvesting.

We will characterize the total amount of correlations with the mutual information, 
\begin{equation}
I(\rho_{AB})=S(\rho_A)+S(\rho_B)-S(\rho_{AB}),
\end{equation}
where $\rho_A=\text{Tr}_{B}(\rho_{AB})$ (and vice versa) is the partial trace of $\rho_{AB}$ with respect to subsystem $B$ ($A$) and $S(\rho)=-\text{Tr}(\rho\log\rho)$ is the von Neumann entropy.

For the state given by \eqref{eq:seconddens}, the partial subsystems are described by the following density matrices
\begin{equation}
\rho_\mu=\begin{pmatrix}
1-\mathcal{L}_{\mu\mu} & 0 \\
0 & \mathcal{L}_{\mu\mu}
\end{pmatrix},
\end{equation}
so, once expanded to leading order in powers of the coupling strength, the mutual information takes the form
\begin{align}
I(\rho_{AB})=&\mathcal{L}_{+} \log (\mathcal{L}_{+})+\mathcal{L}_{-} \log (\mathcal{L}_{-})\notag\\
&-\mathcal{L}_{AA} \log (\mathcal{L}_{AA})-\mathcal{L}_{BB} \log (\mathcal{L}_{BB})+\mathcal{O}(\lambda_\nu^4),
\end{align}
where the quantities $\mathcal{L}_\pm$ are defined by
\begin{equation}
\mathcal{L}_\pm=\frac{1}{2}\left(\mathcal{L}_{AA}+\mathcal{L}_{BB}\pm\sqrt{(\mathcal{L}_{AA}-\mathcal{L}_{BB})^2+4 \left|\mathcal{L}_{AB}\right|^2}\right).
\end{equation}

Note that in this expression the competition between the local ($\mathcal{L}_{\mu\mu}$) and nonlocal ($\mathcal{L}_\pm$) terms is also explicit.

\begin{figure*}
\begin{tabular}{cccc}
\includegraphics[width=0.24\textwidth]{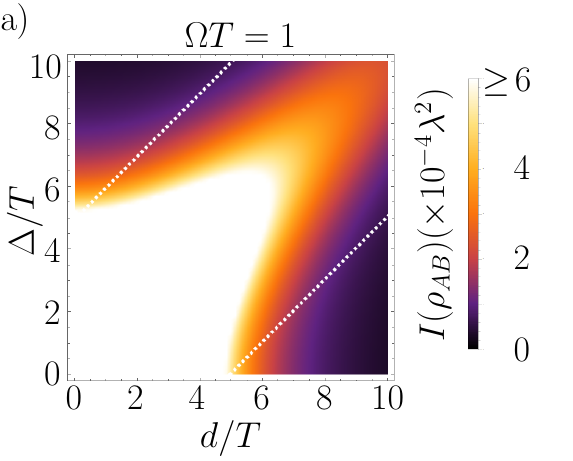}
&\includegraphics[width=0.24\textwidth]{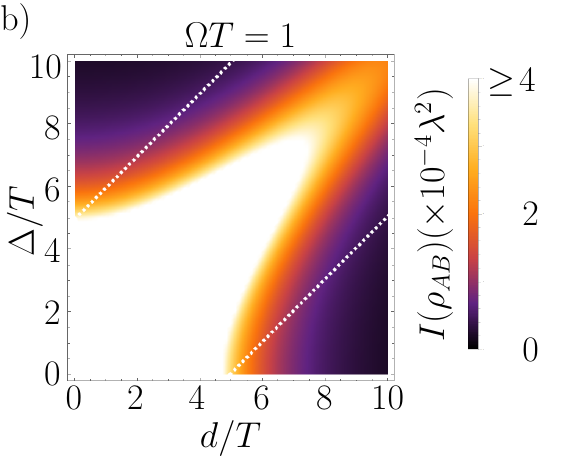}
&\includegraphics[width=0.24\textwidth]{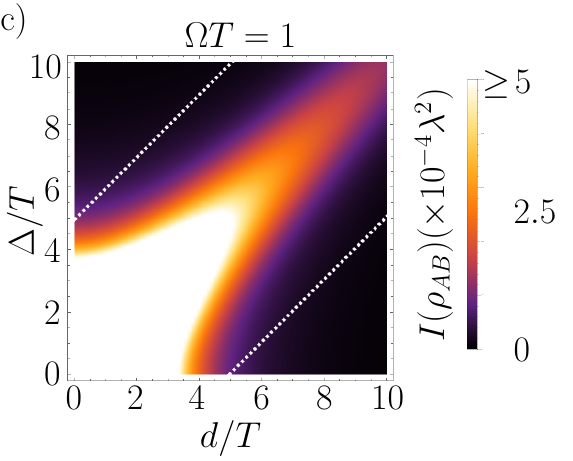}
&\includegraphics[width=0.24\textwidth]{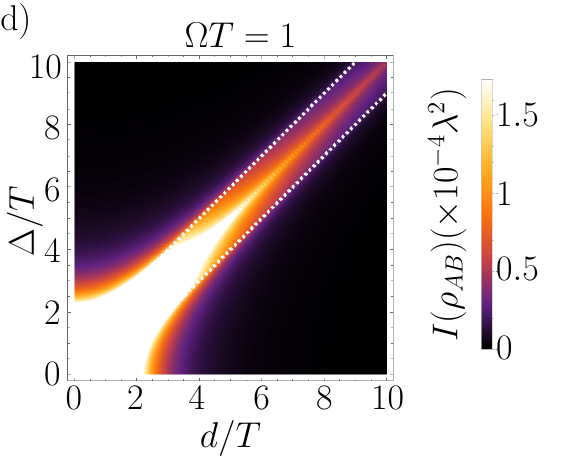}\\
\includegraphics[width=0.24\textwidth]{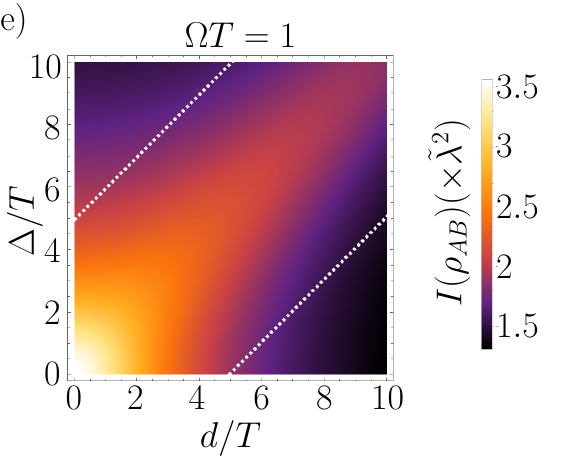}
&\includegraphics[width=0.24\textwidth]{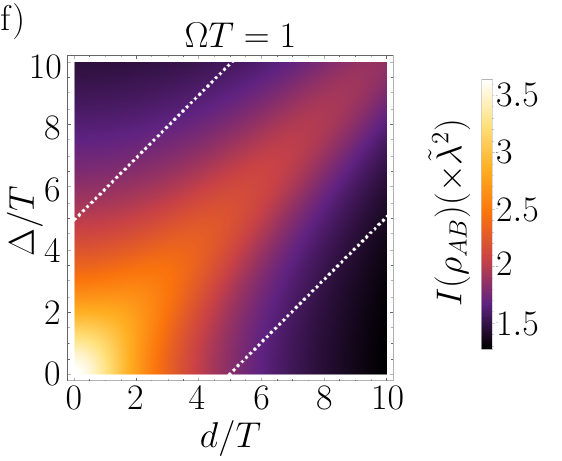}
& \includegraphics[width=0.24\textwidth]{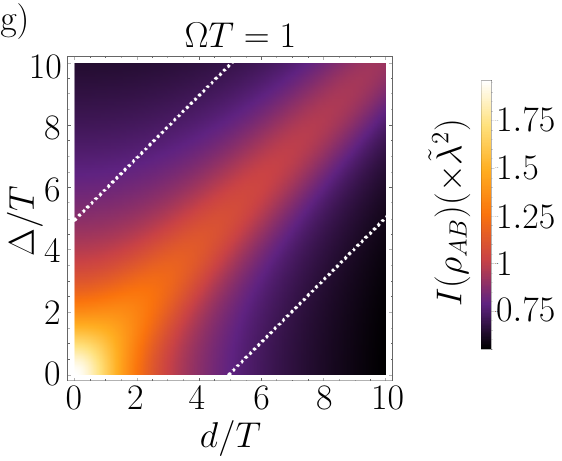}
& \includegraphics[width=0.24\textwidth]{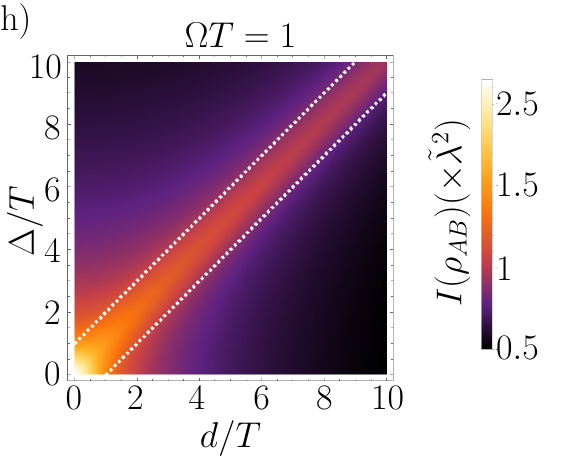}
\end{tabular}
\caption{Mutual Information for all the cases described in sections \ref{3D} and \ref{1D}, in the following order: (a) 3D Gaussian switching and Gaussian spatial profile; (b) 3D Gaussian switching and pointlike detectors; (c) 3D sudden switching and Gaussian spatial profile; (d) 3D sudden switching and near-pointlike detectors ($\sigma/T\ll1$); (e) 1D Gaussian switching and Gaussian spatial profile; (f) 1D Gaussian switching and pointlike detectors; (g) 1D sudden switching and Gaussian spatial profile; (h) 1D sudden switching and pointlike detectors. The graphs show the harvested mutual information as a function of the spatial separation of the detectors and the time delay between their switching functions for $\Omega T=1$. In all plots the dashed lines represent the boundaries of the lightcone. These boundaries are placed at $\Delta=d\pm T$ in the cases of pointlike detectors which are switched suddenly, and at $\Delta=d\pm 7T/\sqrt{2}$ in the rest, which is a reasonable estimation given the discussion in section II, subsection A1. All cases of Gaussian spatial profile have $\sigma/T=1$ except for the near-pointlike detectors (plot d), for which $\sigma/T=0.01$.}
\label{mutinfs}
\end{figure*}

In the same fashion as in section \ref{entharvest}, we show in Fig. \ref{mutinfs} a collection of representative examples of the mutual information for all the cases defined in sections \ref{3D} and \ref{1D}.

As we expected from the discussion in the study of the negativity, the mutual information maximizes inside the lightcone. In the same way as for the negativity, the total correlation between the detectors peaks inside the region of light contact, where direct exchange of real field quanta is possible.

\begin{figure}[h]
\centering
\includegraphics[width=0.48\textwidth]{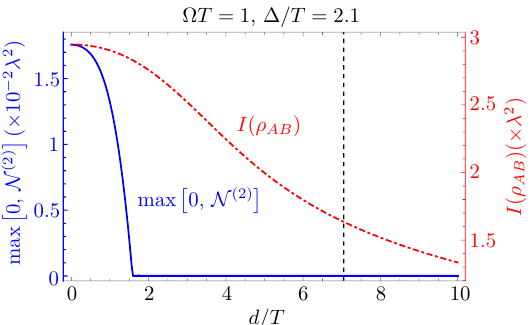}
\caption{Mutual information (dashed red) and negativity (solid blue) to $\mathcal{O}(\lambda_\nu^2)$ for 1+1-dimensional detectors with Gaussian spatial smearing and Gaussian switching. The black, dashed line represents the usual estimation of the effective end of the lightcone $\beta=\gamma+7/\sqrt{2}$.}
\label{classicalcorrelations}
\end{figure}

We first note that, unlike entanglement, harvesting of mutual information is possible for the whole spacetime, even for large temporal and spatial separations between the detectors long after entanglement harvesting is no longer possible (see Fig. \ref{classicalcorrelations}).

Due to this, when the separation between the detectors is large enough, we can identify the main source of harvested correlations as classical correlations present in the field vacuum. Nevertheless, we should not forget that other sources of quantum correlations (quantum discord \cite{BrownDiscord}) may be present, and their study in future works may be of interest.

In Fig. \ref{mutinfs}, we see how the dimensionality of spacetime influences the harvesting of correlations: the gain of mutual information during light contact is more efficient in 3+1 dimensions than in 1+1. Interestingly, for a given value of $\Omega T$ the amount of correlations harvested in the spacelike-separation region is higher in the 1+1-dimensional case.

We also observe the same effect of ``damping and leakage'' that was observed in section \ref{entharvest} when increasing the internal energy gap is observed in the case of correlations: increasing $\Omega T$ decreases the overall value of the mutual information but also makes its support outside the lightcone increase.

\section{Conclusions}

We have performed a detailed study of the phenomenon of entanglement and mutual information harvesting from the vacuum state of a scalar field, using a pair of Unruh-DeWitt particle detectors.

First, we have analyzed the influence on this phenomenon of various aspects, such as the dimensionality of spacetime, the nature of the switching of the detectors (sudden vs. smooth), the detectors' physical size and their internal energy structure.

While the dimensionality of spacetime does not seem to strongly influence entanglement harvesting, the gain of mutual information during light contact is more efficient in 3+1 dimensions than in 1+1 dimensions (e.g, inside an optical fiber). Conversely, the amount of spacelike-correlation harvesting is noticeably larger in the 1+1-dimensional case.

We have made a comparative study of two different kinds of switching: I) smooth Gaussian and II) sudden switching. We have shown that the smooth Gaussian switching is much more efficient than the sudden switching to harvest entanglement from the vacuum in all cases: spacelike, timelike and lightlike.  Remarkably, we have found that it is not possible to harvest spacelike entanglement with sudden switching, neither in 3+1 nor 1+1 dimensions, for the parameter landscape studied (with the exception of pointlike detectors in 1+1 dimensions for a very limited range of detector configurations and small spatial separations very close to the lightcone). This difficulty of harvesting entanglement with abruptly switched detectors contrasts with the case of smooth Gaussian switching, where it is always possible to find detector configurations for which there is significant entanglement harvesting for spacelike separated detectors arbitrarily far away from the lightcone. This result is striking, and runs perhaps contrary to previous conjectures (suggested by results obtained with non-perturbative harmonic oscillator detectors \cite{Brown2012}) that sudden switching may aid in generating entanglement. Our result is especially interesting in the light of the pioneering studies on entanglement harvesting that showed that super-oscillatory switching functions can enhance the detectors' ability to harvest entanglement \cite{Reznik2005}.

However, the fact that Gaussian switching is more efficient than sudden switching to harvest entanglement stems from the following reason: the more sudden the switching is, the stronger the local noise ($\mathcal{L}_{\mu\mu}$) becomes \cite{Louko2008}. The local noise competes with the nonlocal terms ($\mathcal{M}$) which give rise to entanglement. The smooth Gaussian switching strongly attenuates this noise and therefore facilitates that the nonlocal terms dominate over the local noise terms.

As for the size of the detectors, we have shown that if the detectors are smaller than the characteristic interaction timescale, their harvesting abilities do not substantially deviate from the pointlike case. When the size of the detectors becomes comparable to the interaction timescale, their ability to harvest entanglement gets increasingly hindered as their size increases.

The energy gap of the detectors has a strong influence in the extraction of entanglement and correlations from the vacuum, but this influence is radically different for smoothly switched detectors and suddenly switched detectors. For Gaussian switching it is possible to find values of the gap for which there is always spacelike entanglement harvesting. In more detail, increasing the dimensionless parameter $\Omega T$ leads to a ``damping and leakage'' effect, which allows entanglement to be harvested in a broader range of spatiotemporal distances at the cost of harvesting less entanglement. This is not the case for sudden switching, where increasing the energy gap does not necessarily mean an increase in the area where entanglement can be harvested. This is consistent with our previous result that Gaussian switching is generally better to harvest vacuum entanglement.

We have also discussed that harvesting classical correlations (and possibly quantum discord) is generally easier than harvesting entanglement. In particular it is possible to harvest mutual information when we place the two detectors anywhere in the whole spacetime. Even though the amount of harvested correlation quickly decreases with the spatiotemporal distance, it only vanishes in the limit of infinite separation.

Finally, we have analyzed the first subleading-order correction to the negativity. We have seen that the next order contribution to the negativity vanishes generally at smaller spacetime distances than the second-order contribution, and thus going beyond leading order does not reveal new phenomenology. Therefore a leading-order perturbative approximation is generally enough to identify the regimes in which Unruh-DeWitt detectors can harvest quantum entanglement from the field vacuum.

\begin{acknowledgments}

The authors gratefully acknowledge Luis J. Garay for his valuable comments on the final version of this paper. We also thank Petar Simidzija for his inestimable help reviewing the latest versions of this manuscript. The work of E. M.-M. is supported by the National Sciences and Engineering Research Council of Canada through the Discovery program.
\end{acknowledgments}

\appendix

\section{Explicit computation of $G_2(\kappa)$, $\mathcal{L}_{\mu\nu}$ and $\mathcal{M}$ in the 3D, Gaussian switching case.}\label{AppendixGaussian}

Obtaining a closed-form expression of $G_2(\bm\kappa)$ requires a more subtle process than $G_1(\bm\kappa,\tau_\mu)$ due to the nesting of the $\tau_1$, $\tau_2$ integrals. Recall that we start from
\begin{align}
G_2(\bm{\kappa})=&T^2 e^{2\ii\alpha\tau_A}\int_{-\infty}^\infty\!\!\!\text{d}\tau_1\int_{-\infty}^{\tau_1}\!\!\!\text{d}\tau_2\,e^{\ii\alpha(\tau_1+\tau_2)}e^{-\ii|\bm{\kappa}|(\tau_1-\tau_2)}\notag\\
&\times\left(e^{-\left(\tau_1-\gamma\right)^2}e^{-\tau_2^2}+e^{-\left(\tau_2-\gamma\right)^2}e^{-\tau_1^2}\right).
\end{align}

The innermost integral can be evaluated straightforwardly, yielding
\begin{align}
G_2(\bm{\kappa})=&\frac{\sqrt{\pi}}{2}T^2 e^{2\ii\alpha\tau_A}\notag\\
&\times\int_{-\infty}^{\infty}\text{d}\tau_1\,\bigg\{ e^{-\left(\tau_1-\gamma\right)^2-\ii \tau_1 \left(|\bm{\kappa}|-\alpha \right)-\frac{1}{4}\left(|\bm{\kappa}|+\alpha\right)^2}\notag\\
&\times\left[1+\text{erf}\left(\tau_1-\frac{1}{2} \ii(|\bm{\kappa}|+\alpha )\right)\right]\notag\\
&+e^{-\tau_1^2-\ii \left[\tau_1 (|\bm{\kappa}|-\alpha )-\gamma(|\bm{\kappa}|+\alpha )\right]-\frac{1}{4} \left(|\bm{\kappa}|+\alpha \right)^2}\notag\\
&\times\left.\left[1+\text{erf}\left(\tau_1-\gamma -\frac{1}{2}\ii(|\bm{\kappa}|+\alpha )\right)\right]\right\}\label{G2firstint}.
\end{align}

Notice that \eqref{G2firstint} has four summands. Two of them are combinations of exponential and Gaussian functions, whereas the other two are of the form
\begin{equation}
I(a,b)=\int_{-\infty}^\infty\text{d}y\,e^{-a^2-\ii b y -y^2}\text{erf}\left(y-\ii a\right).\label{interf}
\end{equation}

Partial differentiation under the integral sign of the previous expression with respect to the parameter $a$ yields
\begin{align}
\frac{\partial}{\partial a}I(a,b)=&\int_{-\infty}^\infty\text{d}y\,\left[-2a e^{-a^2-\ii b y -y^2}\text{erf}\left(y-\ii a\right)\right.\notag\\
&\left.-\ii\frac{2}{\sqrt{\pi}}e^{-a^2+\left(a+\ii y\right)^2-\ii b y-y^2}\right]\notag\\
=&-2\,a\,I(a,b)-\ii\sqrt{2}e^{-\frac{1}{8}\left(b-2a\right)^2}.\label{diffeq}
\end{align}

This is a linear, first-order, nonhomogeneous diferential equation, which can be solved, for instance, via variation of constants. The solution to the homogeneous equation is simply
\begin{equation}
I_H(a,b)=Ce^{-a^2},
\end{equation}
and now, allowing the constant to be a function of $a$ and $b$, and inserting $I_H(a,b)$ in Eq. \eqref{diffeq}, we obtain
\begin{align}
&\frac{\partial}{\partial a}C(a,b)e^{-a^2}=-\ii\sqrt{2}e^{-\frac{1}{8}\left(b-2a\right)^2}\notag\\
&C(a,b)=-\ii\sqrt{\pi}e^{-\frac{b^2}{4}}\text{erfi}\left(\frac{2a+b}{2\sqrt{2}}\right).
\end{align}

So the closed form on the integral $I(a,b)$ is
\begin{equation}
I(a,b)=-\ii\sqrt{\pi}e^{-a^2-\frac{b^2}{4}}\text{erfi}\left(\frac{a+\frac{b}{2}}{\sqrt{2}}\right).
\end{equation}

Therefore, using this in \eqref{G2firstint} we get \eqref{bothgausstime},
\begin{align}
G_2(\bm\kappa)=&T^2\frac{\sqrt{\pi}}{2} e^{2\ii\alpha\tau_A}\bigg[\sqrt{\pi}e^{-\frac{1}{2}\left(\alpha^2+\left|\bm\kappa\right|^2\right)+\ii\gamma(\alpha-|\bm\kappa|)}\notag\\
&+e^{-\gamma^2}\,I\Big(\frac{1}{2}(|\bm\kappa|+\alpha),|\bm\kappa|-\alpha+2\ii\gamma\Big)\bigg]\notag\\
&+T^2\frac{\sqrt{\pi}}{2} e^{2\ii\alpha\tau_A}\bigg[\sqrt{\pi}e^{-\frac{1}{2}\left(\alpha^2+\left|\bm\kappa\right|^2\right)+\ii\gamma(\alpha+|\bm\kappa|)}\notag\\
&+e^{-\gamma^2}\,I\Big(\!-\!\ii\gamma\!+\!\frac{1}{2}(|\bm\kappa|+\alpha),|\bm\kappa|\!-\!\alpha\Big)\bigg].
\end{align}

Now that we have closed expressions for $G_1(\bm\kappa,\tau_\mu)$ and $G_2(\bm\kappa)$ we can compute $\mathcal{L}_{\mu\nu}$ and $\mathcal{M}$. Recall that, from \eqref{Lmunu} and \eqref{M},
\begin{align}
\mathcal{L}_{\mu\mu}\!=&\lambda^2\!\!\int\!\!\text{d}^3\bm{k}\,\frac{e^{-\frac{1}{2}\bm{k}^2\sigma^2}}{\left(2\pi\right)^32|\bm{k}|}G_1(\bm\kappa,\tau_\mu)G_1^*(\bm\kappa,\tau_\mu),\\
\mathcal{L}_{AB}\!=&\lambda^2\!\!\int\!\!\text{d}^3\bm{k}\,\frac{e^{-\ii\bm{k}\cdot\Delta\bm{x}}e^{-\frac{1}{2}\bm{k}^2\sigma^2}}{\left(2\pi\right)^32|\bm{k}|}G_1(\bm\kappa,0)G_1^*(\bm\kappa,\gamma),\\
|\mathcal{M}|=&\lambda^2\,\left|\int\text{d}^3\bm{k}\,e^{\ii\bm{k}\cdot(\bm{x}_{A}-\bm{x}_{B})}\frac{e^{-\frac{1}{2}\bm{k}^2\sigma^2}}{\left(2\pi\right)^32|\bm{k}|}G_2(\bm{\kappa})\right|.
\end{align}

Writing these integrals in terms of the dimensionless variable \mbox{$\bm\kappa=\bm k T$}, using spherical coordinates and choosing the $z$-axis in the direction of $\bm{\kappa}$ we obtain:
\begin{align}
\mathcal{L}_{\mu\mu}=&\frac{\lambda^2}{4\pi^2T^2}\int_0^\infty\!\!\!\!\text{d}|\bm{\kappa}|\,|\bm\kappa|\,e^{-\frac{1}{2}\bm{\kappa}^2\delta^2}\left|G_1(\bm\kappa,\tau_\mu)\right|^2\notag\\
=&\frac{\lambda ^2 e^{-\frac{1}{2}\alpha^2}}{8 \pi  \left(1+\delta ^2\right)}\!\left[2-\frac{\sqrt{2 \pi }\, \alpha \,e^{\frac{\alpha ^2}{2 \left(1+\delta ^2\right)}} \text{erfc}\left(\frac{\alpha }{\sqrt{2} \sqrt{1+\delta ^2}}\right)}{\sqrt{1+\delta ^2}}\right]\!,
\end{align}
which gives the values for $\mathcal{L}_{AA}$ and $\mathcal{L}_{BB}$,

\begin{align}
\mathcal{L}_{AB}=&\frac{\lambda^2}{8\pi^2}\int_0^\infty\text{d}|\bm{\kappa}|\,\frac{|\bm{\kappa}|^2}{T^2}\int_{-1}^1\text{d}(\cos\theta)\notag\\
&\frac{e^{-\ii|\bm{\kappa}|\beta\cos\theta}}{|\bm{\kappa}|}e^{-\frac{1}{2}\bm{\kappa}^2\delta^2}G_1(\bm\kappa,0)G_1^*(\bm\kappa,\gamma)\notag\\
=&\frac{\lambda^2}{4\pi^2T^2\beta}\int_0^\infty\!\!\!\!\!\text{d}|\bm{\kappa}|\,\sin(|\bm{\kappa}|\beta)e^{-\frac{1}{2}\bm{\kappa}^2\delta^2}G_1(\kappa,0)G_1^*(\kappa,\gamma)\notag\\
=&\frac{\ii\lambda^2 e^{-\frac{1}{2} \alpha^2}e^{-\ii\alpha\gamma}}{8 \sqrt{2\pi} \beta \sqrt{1+\delta^2}}\notag\\
&\times\left\{e^{-\frac{\left(\beta+\gamma-\ii \alpha \right)^2}{2 \left(1+\delta^2\right)}} \text{erfc}\left(\ii\frac{\beta+\gamma-\ii \alpha }{\sqrt{2} \sqrt{1+\delta^2}}\right)\right.\notag\\
&\left.- e^{-\frac{\left(\beta-\gamma+\ii \alpha \right)^2}{2 \left(1+\delta^2\right)}} \text{erfc}\left(-\ii\frac{\beta-\gamma+\ii\alpha }{\sqrt{2} \sqrt{1+\delta^2}}\right)\right\},
\end{align}
being $\mathcal{L}_{BA}=\mathcal{L}_{AB}^*$, and, finally,
\begin{align}
|\mathcal{M}|=&\frac{\lambda^2}{4\pi^2\beta}\Bigg|\int_0^{\infty}\text{d}|\bm{\kappa}|\,\sin (\beta  |\bm{\kappa}| )\,e^{-\frac{1}{2} \delta ^2 \bm{\kappa} ^2}G_2(\bm{\kappa})\Bigg|\notag\\
=&\frac{\lambda^2e^{-\frac{1}{2}\alpha ^2}}{8\pi\beta}\Bigg|\int_0^{\infty}\text{d}|\bm{\kappa}|\,\sin (\beta  |\bm{\kappa}| )\,e^{-\frac{1}{2} \left(1+\delta ^2\right) \bm{\kappa} ^2}\notag\\
&\!\!\!\!\!\!\!\times\!\bigg[e^{-\ii\left| \bm\kappa \right|\gamma}\,\text{erfc}\!\left(\!\frac{-\gamma \!+\!\ii |\bm\kappa|}{\sqrt{2}}\!\right)\!+ \!e^{\ii |\bm\kappa|\gamma}\,\text{erfc}\!\left(\frac{\gamma +\ii |\bm\kappa| }{\sqrt{2}}\right)\bigg]\Bigg|.
\end{align}

\section{Explicit computation of $S_2(\kappa)$}\label{AppendixSudden}

To find simple expressions for $S_2(\bm\kappa)$, we will separate the cases in which the switching functions do and do not overlap. In the nonoverlapping case, let us assume without loss of generality that detector A is switched off before detector B is switched on. In this case, the first contribution of \eqref{S2} is trivially zero, and in the second one the time integrals decouple. Substituting the dimensionless variables $\tau_i=t_i/T$ we get
\begin{align}
{S_2}_\text{non}(\bm\kappa)=&T^2\int_{\tau_B^\text{on}}^{\tau_B^\text{off}}\text{d}\tau_1\int_{\tau_A^\text{on}}^{\tau_A^\text{off}}\text{d}\tau_2\,e^{\ii\alpha(\tau_1+\tau_2)}e^{-\ii|\bm\kappa|(\tau_1-\tau_2)}\notag\\
=&\frac{\left(e^{\ii (\alpha -\left| \bm{\kappa} \right| ) (1+\gamma)}-e^{\ii (\alpha -\left| \bm{\kappa} \right| ) \gamma}\right)\left(e^{\ii \left(\alpha +\left| \bm{\kappa} \right| \right)}-1\right)}{\left(\bm{\kappa}^2-\alpha ^2\right)T^{-2}}\label{S2nonover}
\end{align}
where $\tau_\mu^{\text{on,off}}=T_\mu^{\text{on,off}}/T$.

When the switching functions overlap, a more careful analysis is required. We assume without loss of generality that detector A is switched on simultaneously, or before detector B. For illustration let us further assume that detector B is switched off simultaneously or after detector A. The total interaction time interval $\left[T_A^\text{on},T_B^\text{off}\right]$ can be then subdivided into different regions:
\begin{itemize}
\item Regions of no overlap: $\left[T_A^\text{on},T_B^\text{on}\right]\cup\left[T_A^\text{off},T_B^\text{off}\right]$.
\item Region of overlap: $\left(T_B^\text{on},T_A^\text{off}\right)$.
\end{itemize}

Thus we will have three different contributions,
\begin{equation}
{S_2}(\bm\kappa)={S_2}^{\left[T_A^\text{on},T_B^\text{on}\right]}(\bm\kappa)+{S_2}^{\left[T_A^\text{off},T_B^\text{off}\right]}(\bm\kappa)+{S_2}^{\left(T_B^\text{on},T_A^\text{off}\right)}(\bm\kappa).
\end{equation}

In the nonoverlapping regions we get a result analogous to \eqref{S2nonover} ---the first contribution is zero and the integrals decouple in the second--, so we obtain
\begin{align}
{S_2}&^{\left[T_A^\text{on},T_B^\text{on}\right]}(\bm\kappa)=T^2\!\!\int_{\tau_B^\text{on}}^{\tau_B^\text{off}}\!\!\!\!\text{d}\tau_1\int_{\tau_A^\text{on}}^{\tau_B^\text{on}}\!\!\!\!\text{d}\tau_2\,e^{\ii\alpha(\tau_1+\tau_2)}e^{-\ii|\bm\kappa|(\tau_1-\tau_2)}\notag\\
=&\frac{T^2e^{\ii\gamma(\alpha-\left|\bm\kappa\right|)}}{\alpha ^2-\bm\kappa^2}\notag\\
&\qquad\quad\!\!\times\left(e^{\ii(\alpha-\left| \bm\kappa\right|)}\!-\!e^{\ii\left[\alpha(\gamma+1)+\left| \bm\kappa\right|  (\gamma-1)\right]}\!+\!e^{ \ii \gamma (\alpha+\left|\bm\kappa\right|) }\!-\!1\right)\!,\notag\\
{S_2}&^{\left[T_A^\text{off},T_B^\text{off}\right]}(\bm\kappa)=T^2\!\!\int_{\tau_A^\text{off}}^{\tau_B^\text{off}}\!\!\!\!\text{d}\tau_1\int_{\tau_B^\text{on}}^{\tau_A^\text{off}}\!\!\!\!\text{d}\tau_2\,e^{\ii\alpha(\tau_1+\tau_2)}e^{-\ii|\bm\kappa|(\tau_1-\tau_2)}\notag\\
=&\frac{T^2e^{\ii \alpha(\gamma+1)}}{\alpha ^2-\bm\kappa ^2}\!\!\left(\!e^{\ii\alpha(\!1-\gamma)}\!-\!e^{\ii(\gamma-1)|\bm\kappa|}\!-\!e^{\ii (\alpha-\gamma|\bm\kappa|)}\!+\!e^{\ii \left(\alpha  \gamma -\left| \bm\kappa \right| \right)}\!\right)\!.
\end{align}

In the overlapping region both contributions of \eqref{S2} are equal and nonzero, so we can write
\begin{align}
{S_2}&^{\left(T_B^\text{on},T_A^\text{off}\right)}(\bm\kappa)=2T^2\!\!\int_{\tau_B^\text{on}}^{\tau_A^\text{off}}\!\!\!\!\text{d}\tau_1\!\int_{\tau_B^\text{on}}^{\tau_1}\!\!\!\!\text{d}\tau_2\,e^{\ii\alpha(\tau_1+\tau_2)}e^{-\ii|\bm\kappa|(\tau_1-\tau_2)}\notag\\
=&2T^2\left[\frac{e^{\ii \left[\left| \bm\kappa\right|  (\gamma-1)+\alpha  (\gamma+1)\right]}-e^{2 \ii \gamma \alpha }}{\alpha ^2-\bm\kappa^2}+\frac{e^{2 \ii \gamma \alpha }-e^{2 \ii\alpha }}{2\alpha  (\left| \bm\kappa\right| +\alpha )}\right].
\end{align}

Adding all three contributions, we obtain \eqref{S2overint}.

\section{The fourth-order term $\Xi_3$}\label{computeXi}

The four components $\xi_i(\bm\kappa,\bm\eta),\,i=1,\dots,4$ of $\Xi_3$ are
\begin{align}
\xi_1(\bm k,\bm q)=&\lambda_A^2\lambda_B^2\frac{[\tilde{F}(\bm k)]^2[\tilde{F}(\bm q)]^2}{4|\bm k||\bm q|}\notag\\
&\!\!\!\!\!\!\!\!\!\!\!\times\int_{-\infty}^{\infty}\!\!\!\!\!\!\!\text{d}t_1\!\!\int_{-\infty}^{t_1}\!\!\!\!\!\!\!\text{d}t_2\!\!\int_{-\infty}^{\infty}\!\!\!\!\!\!\!\text{d}t_1'\!\!\int_{-\infty}^{t_1'}\!\!\!\!\!\!\!\text{d}t_2'\,\chi_A(t_1)\chi_B(t_2)\chi_A(t_1')\chi_B(t_2')\notag\\
&\!\!\!\!\!\!\!\!\!\!\!\times e^{\ii(\Omega_{A}t_1+\Omega_{B}t_2-\Omega_{A}t_1'-\Omega_{B}t_2')}\notag\\
&\!\!\!\!\!\!\!\!\!\!\!\times\Big[e^{\ii|\bm{k}|(t_1-t_2')}e^{\ii|\bm{q}|(t_2-t_1')}e^{-\ii(\bm{k}-\bm{q})\cdot(\bm{x}_{A}-\bm{x}_{B})}\notag\\
&\!\!\!\!\!\!\!\!\!\!\!+e^{\ii|\bm{k}|(t_2-t_1)}e^{\ii|\bm{q}|(t_1'-t_2')}e^{\ii(\bm{k}-\bm{q})\cdot(\bm{x}_{A}-\bm{x}_{B})}\notag\\
&\!\!\!\!\!\!\!\!\!\!\!+e^{\ii|\bm{k}|(t_1-t_1')}e^{\ii|\bm{q}|(t_2-t_2')}\Big],
\end{align}
\begin{align}
\xi_2(\bm k,\bm q)=&\lambda_A^2\lambda_B^2\frac{[\tilde{F}(\bm k)]^2[\tilde{F}(\bm q)]^2}{4|\bm k||\bm q|}\notag\\
&\!\!\!\!\!\!\!\!\!\!\!\times\int_{-\infty}^{\infty}\!\!\!\!\!\!\!\text{d}t_1\!\!\int_{-\infty}^{t_1}\!\!\!\!\!\!\!\text{d}t_2\!\!\int_{-\infty}^{\infty}\!\!\!\!\!\!\!\text{d}t_1'\!\!\int_{-\infty}^{t_1'}\!\!\!\!\!\!\!\text{d}t_2'\,\chi_A(t_1)\chi_B(t_2)\chi_B(t_1')\chi_A(t_2')\notag\\
&\!\!\!\!\!\!\!\!\!\!\!\times e^{\ii(\Omega_{A}t_1+\Omega_{B}t_2-\Omega_{B}t_1'-\Omega_{A}t_2')}\notag\\
&\!\!\!\!\!\!\!\!\!\!\!\times\Big[e^{\ii|\bm{k}|(t_1-t_1')}e^{\ii|\bm{q}|(t_2-t_2')}e^{-\ii(\bm{k}-\bm{q})\cdot(\bm{x}_{A}-\bm{x}_{B})}\notag\\
&\!\!\!\!\!\!\!\!\!\!\!+e^{\ii|\bm{k}|(t_2-t_1)}e^{\ii|\bm{q}|(t_1'-t_2')}e^{\ii(\bm{k}+\bm{q})\cdot(\bm{x}_{A}-\bm{x}_{B})}\notag\\
&\!\!\!\!\!\!\!\!\!\!\!+e^{\ii|\bm{k}|(t_1-t_2')}e^{\ii|\bm{q}|(t_2-t_1')}\Big],
\end{align}
\begin{align}
\xi_3(\bm k,\bm q)=&\lambda_A^2\lambda_B^2\frac{[\tilde{F}(\bm k)]^2[\tilde{F}(\bm q)]^2}{4|\bm k||\bm q|}\notag\\
&\!\!\!\!\!\!\!\!\!\!\!\times\int_{-\infty}^{\infty}\!\!\!\!\!\!\!\text{d}t_1\!\!\int_{-\infty}^{t_1}\!\!\!\!\!\!\!\text{d}t_2\!\!\int_{-\infty}^{\infty}\!\!\!\!\!\!\!\text{d}t_1'\!\!\int_{-\infty}^{t_1'}\!\!\!\!\!\!\!\text{d}t_2'\,\chi_B(t_1)\chi_A(t_2)\chi_A(t_1')\chi_B(t_2')\notag\\
&\!\!\!\!\!\!\!\!\!\!\!\times e^{\ii(\Omega_{B}t_1+\Omega_{A}t_2-\Omega_{A}t_1'-\Omega_{B}t_2')}\notag\\
&\!\!\!\!\!\!\!\!\!\!\!\times\Big[e^{\ii|\bm{k}|(t_1-t_1')}e^{\ii|\bm{q}|(t_2-t_2')}e^{\ii(\bm{k}-\bm{q})\cdot(\bm{x}_{A}-\bm{x}_{B})}\notag\\
&\!\!\!\!\!\!\!\!\!\!\!+e^{\ii|\bm{k}|(t_2-t_1)}e^{\ii|\bm{q}|(t_1'-t_2')}e^{-\ii(\bm{k}+\bm{q})\cdot(\bm{x}_{A}-\bm{x}_{B})}\notag\\
&\!\!\!\!\!\!\!\!\!\!\!+e^{\ii|\bm{k}|(t_1-t_2')}e^{\ii|\bm{q}|(t_2-t_1')}\Big],
\end{align}
\begin{align}
\xi_4(\bm k,\bm q)=&\lambda_A^2\lambda_B^2\frac{[\tilde{F}(\bm k)]^2[\tilde{F}(\bm q)]^2}{4|\bm k||\bm q|}\notag\\
&\!\!\!\!\!\!\!\!\!\!\!\times\int_{-\infty}^{\infty}\!\!\!\!\!\!\!\text{d}t_1\!\!\int_{-\infty}^{t_1}\!\!\!\!\!\!\!\text{d}t_2\!\!\int_{-\infty}^{\infty}\!\!\!\!\!\!\!\text{d}t_1'\!\!\int_{-\infty}^{t_1'}\!\!\!\!\!\!\!\text{d}t_2'\,\chi_B(t_1)\chi_A(t_2)\chi_B(t_1')\chi_A(t_2')\notag\\
&\!\!\!\!\!\!\!\!\!\!\!\times e^{\ii(\Omega_{B}t_1+\Omega_{A}t_2-\Omega_{B}t_1'-\Omega_{A}t_2')}\notag\\
&\!\!\!\!\!\!\!\!\!\!\!\times\Big[e^{\ii|\bm{k}|(t_2-t_1)}e^{\ii|\bm{q}|(t_1'-t_2')}e^{-\ii(\bm{k}-\bm{q})\cdot(\bm{x}_{A}-\bm{x}_{B})}\notag\\
&\!\!\!\!\!\!\!\!\!\!\!+e^{\ii|\bm{k}|(t_1-t_2')}e^{\ii|\bm{q}|(t_2-t_1')}e^{\ii(\bm{k}-\bm{q})\cdot(\bm{x}_{A}-\bm{x}_{B})}\notag\\
&\!\!\!\!\!\!\!\!\!\!\!+e^{\ii|\bm{k}|(t_1-t_1')}e^{\ii|\bm{q}|(t_2-t_2')}\Big].
\end{align}

To evaluate these integrals, let us consider the simpler case where the detectors are identical and the detectors' switching functions have supports that are far apart. In this case the only nonzero contribution to $\Xi_3$ is $\xi_4$, which can, under these assumptions, be split into three parts that we denote as:
\begin{align}
{\xi_4}_a(\bm k,\bm q)=&\lambda^4\frac{[\tilde{F}(\bm k)]^2[\tilde{F}(\bm q)]^2}{4|\bm k||\bm q|}e^{-\ii(\bm\kappa-\bm\eta)\cdot(\bm\beta_A-\bm\beta_B)}R_1(\bm\kappa,\bm\eta),
\end{align}
\begin{align}
{\xi_4}_b(\bm k,\bm q)=&\lambda^4\frac{[\tilde{F}(\bm k)]^2[\tilde{F}(\bm q)]^2}{4|\bm k||\bm q|}e^{\ii(\bm\kappa-\bm\eta)\cdot(\bm\beta_A-\bm\beta_B)}\!R_2(\bm\kappa,\bm\eta),
\end{align}
\begin{align}
{\xi_4}_c(\bm k,\bm q)=&\lambda^4\frac{[\tilde{F}(\bm k)]^2[\tilde{F}(\bm q)]^2}{4|\bm k||\bm q|}R_3(\bm\kappa,\bm\eta),
\end{align}
where $\bm\kappa=\bm k T$ and $\bm\eta=\bm q T$ are dimensionless momenta and the functions $R_i(\bm\kappa,\bm\eta)$ are given by
\begin{align}
R_1(\bm\kappa,\bm\eta)=&\int_{-\infty}^{\infty}\text{d}\tau_1\int_{-\infty}^{\infty}\text{d}\tau_2\int_{-\infty}^{\infty}\text{d}\tau_1'\int_{-\infty}^{\infty}\text{d}\tau_2'\notag\\
&\chi_A(\tau_1)\chi_B(\tau_2)\chi_A(\tau_1')\chi_B(\tau_2')\notag\\
&\times e^{\ii\alpha(\tau_1+\tau_2-\tau_1'-\tau_2')}e^{\ii|\bm{\kappa}|(\tau_2-\tau_1)}e^{\ii|\bm{\eta}|(\tau_1'-\tau_2')},
\end{align}
\begin{align}
R_2(\bm\kappa,\bm\eta)=&\int_{-\infty}^{\infty}\text{d}\tau_1\int_{-\infty}^{\infty}\text{d}\tau_2\int_{-\infty}^{\infty}\text{d}\tau_1'\int_{-\infty}^{\infty}\text{d}\tau_2'\notag\\
&\chi_A(\tau_1)\chi_B(\tau_2)\chi_A(\tau_1')\chi_B(\tau_2')\notag\\
&\times e^{\ii\alpha(\tau_1+\tau_2-\tau_1'-\tau_2')}e^{\ii|\bm{\kappa}|(\tau_1-\tau_2')}e^{\ii|\bm{\eta}|(\tau_2-\tau_1')},
\end{align}
\begin{align}
R_3(\bm\kappa,\bm\eta)=&\int_{-\infty}^{\infty}\text{d}\tau_1\int_{-\infty}^{\infty}\text{d}\tau_2\int_{-\infty}^{\infty}\text{d}\tau_1'\int_{-\infty}^{\infty}\text{d}\tau_2'\notag\\[4mm]
&\chi_A(\tau_1)\chi_B(\tau_2)\chi_A(\tau_1')\chi_B(\tau_2')\notag\\[3mm]
&\times e^{\ii\alpha(\tau_1+\tau_2-\tau_1'-\tau_2')}e^{\ii|\bm{\kappa}|(\tau_1-\tau_1')}e^{\ii|\bm{\eta}|(\tau_2-\tau_2')}.\notag\\
\end{align}

Thus $\Xi_3$ can be written as
\begin{align}
\Xi_3=&\left(4\pi\right)^2T^2\lambda^4\int_0^{\infty}\text{d}|\bm\kappa|\int_0^{\infty}\text{d}|\bm\eta|\,[\tilde{F}(\bm k)]^2[\tilde{F}(\bm q)]^2\notag\\
&\times\bigg\{\frac{\sin(\beta|\bm\kappa|)\sin(\beta|\bm\eta|)}{\beta^2}\left[R_1(\bm\kappa,\bm\eta)+R_2(\bm\kappa,\bm\eta)\right]\notag\\
&+|\bm\kappa||\bm\eta|R_3(\bm\kappa,\bm\eta)\bigg\}.\label{Xigeneral}
\end{align}
In the case when the detectors' switching functions are two Gaussians far apart enough so that their mutual overlap can be neglected, all the previous integrals admit closed-form expressions. Using \eqref{smear} and \eqref{gauss} we get to the final expression
\begin{widetext}
\begin{align}
\Xi_3=&\frac{\lambda ^4 e^{-\frac{\alpha ^2 \left(1+\delta^2\right)+\beta ^2+\gamma ^2}{1+\delta ^2}} }{128 \pi  \beta ^2 \left(1+\delta ^2\right) }\left\{\!e^{-\frac{(\beta +\gamma )^2}{2 \left(1+\delta ^2\right)}}\!\left[\!1\!+\!\ii\,\text{erfi}\!\left(\!\frac{\beta +\gamma }{\sqrt{2} \sqrt{1+\delta^2}}\!\right)\!\right]\!-\!e^{-\frac{(\beta -\gamma )^2}{2 \left(1+\delta^2\right)}}\!\left[\!1\! -\!\ii \,\text{sign}(\beta\! -\!\gamma ) \text{erfi}\!\left(\!\frac{\left| \beta -\gamma \right| }{\sqrt{2} \sqrt{1+\delta^2}}\!\right)\!\right]\!\right\}\notag\\
&\times\left\{\!e^{\frac{(\beta -\gamma )^2}{2 \left(1+\delta^2\right)}}\!\left[\!1\!-\!\ii \,\text{erfi}\!\left(\!\frac{\beta +\gamma }{\sqrt{2} \sqrt{1+\delta^2}}\!\right)\!\right]\!-\!e^{\frac{(\beta +\gamma )^2}{2 \left(1+\delta^2\right)}}\!\left[\!1\!+\!\ii\,\text{sign}(\beta\! -\!\gamma )\text{erfi}\!\left(\!\frac{\left| \beta -\gamma \right| }{\sqrt{2} \sqrt{1+\delta^2}}\!\right)\!\right]\!\right\}\notag\\
&-\frac{ \lambda ^4e^{-\alpha ^2}}{128 \pi  \beta ^2 \left(1+\delta^2\right)}\left[e^{\frac{[\alpha -\ii (\beta -\gamma )]^2}{2 \left(1+\delta^2\right)}}\text{erfc}\left(\frac{\alpha -\ii (\beta -\gamma )}{\sqrt{2} \sqrt{1+\delta^2}}\right)-e^{\frac{[\alpha +\ii (\beta +\gamma )]^2}{2 \left(1+\delta^2\right)}}\text{erfc}\left(\frac{\alpha +\ii (\beta +\gamma )}{\sqrt{2} \sqrt{1+\delta^2}}\right)\right]\notag\\
&\times\left[e^{\frac{[\alpha -\ii (\beta +\gamma )]^2}{2 \left(1+\delta^2\right)}}\text{erfc}\left(\frac{\alpha -\ii (\beta +\gamma )}{\sqrt{2} \sqrt{1+\delta^2}}\right)-e^{\frac{[\alpha +\ii (\beta -\gamma )]^2}{2 \left(1+\delta^2\right)}} \text{erfc}\left(\frac{\alpha +\ii (\beta -\gamma )}{\sqrt{2} \sqrt{1+\delta^2}}\right)\right]\notag\\
&+\frac{ \lambda ^4e^{-\alpha ^2}}{64 \pi ^2 \left(1+\delta ^2\right)^{3}}\left[2 \sqrt{1+\delta ^2}-\sqrt{2 \pi } \alpha  e^{\frac{\alpha ^2}{2(1+ \delta ^2)}} \text{erfc}\left(\frac{\alpha }{\sqrt{2} \sqrt{1+\delta ^2}}\right)\right]^2.
\end{align}

In the case of nonoverlapping sudden switchings, the different $R_i(\bm\kappa,\bm\eta)$ read
\begin{align}
R_1(\bm\kappa,\bm\eta)=&\frac{e^{-\ii [2 \alpha -(\gamma-1)  \left| \bm\eta \right| +(2 \gamma+1)  \left| \bm\kappa \right| ]}}{\left(\alpha ^2-\left| \bm\eta \right| ^2\right) (\alpha^2 -\left| \bm\kappa \right|^2 )}\left(-e^{\ii \alpha }+e^{\ii (2 \alpha +\left| \bm\eta \right| )}-e^{\ii (\alpha +2 \left| \bm\eta \right| )}+e^{\ii \left| \bm\eta \right| }\right)\left(e^{\ii (\alpha +\left| \bm\kappa \right| )}-1\right) \left(e^{\ii (\alpha +\gamma  \left| \bm\kappa \right| )}-e^{\ii (\gamma +1) \left| \bm\kappa \right| }\right),\notag\\
R_2(\bm\kappa,\bm\eta)=&\frac{e^{-\ii [2 \alpha +(\gamma +1) \left| \bm\eta \right| -(\gamma-1)  \left| \bm\kappa \right| ]}}{(\alpha +\left| \bm\eta \right| )^2 (\alpha +\left| \bm\kappa \right| )^2}\left(e^{\ii (\alpha +\left| \bm\eta \right| )}-1\right)^2\left(e^{\ii (\alpha +\left| \bm\kappa \right| )}-1\right)^2 ,\notag\\
R_3(\bm\kappa,\bm\eta)=&\frac{16 \sin ^2\left(\frac{\alpha +\left| \bm\eta \right| }{2}\right) \sin ^2\left(\frac{\alpha +\left| \bm\kappa \right| }{2}\right)}{(\alpha +\left| \bm\eta \right| )^2 (\alpha +\left| \bm\kappa \right| )^2},
\end{align}
and $\Xi_3$ will be given by \eqref{Xigeneral}, upon substitution of the $R_i(\bm\kappa,\bm\eta)$ integrals.
\end{widetext}
\bibliography{references}

\end{document}